\documentclass[aps,prb,twocolumn,showpacs,superscriptaddress]{revtex4-1}
\usepackage{amsmath}
\usepackage{graphicx}
\usepackage{hyperref}
\usepackage{color}
\usepackage{subfigure}
\def\Tr{\text{Tr}}
\def\stauf{\{(s_1,\tau_1),\cdots, (s_{i},\tau_{i}),\cdots\}}
\def\stauprimef{\{(s_1,\tau_1),\cdots, (-s_{i},\tau_{i}),\cdots\}}
\def\stau{\{ s_{i},\tau_{i}\}} 
\begin{document}

\title{Submatrix updates for the Continuous-Time Auxiliary Field algorithm}
\author{Emanuel Gull}
\affiliation{Department of Physics, Columbia University, New York, NY 10027, USA}
\author{Peter Staar}
\affiliation{Institut f\"{u}r Theoretische Physik, ETH Z\"{u}rich, 8093 Z\"{u}rich, Switzerland}
\author{Sebastian Fuchs}
\affiliation{Institut f\"{u}r Theoretische Physik, Georg-August-Universit\"{a}t G\"{o}ttingen, 37077 G\"{o}ttingen, Germany}
\author{Phani Nukala}
\affiliation{Computer Science and Mathematics Division, Oak Ridge National Laboratory, Oak Ridge, TN 37831-6164, USA}
\author{Michael S. Summers}
\affiliation{Computer Science and Mathematics Division, Oak Ridge National Laboratory, Oak Ridge, TN 37831-6164, USA}
\author{Thomas Pruschke}
\affiliation{Institut f\"{u}r Theoretische Physik, Georg-August-Universit\"{a}t G\"{o}ttingen, 37077 G\"{o}ttingen, Germany}
\author{Thomas C. Schulthess}
\affiliation{Institut f\"{u}r Theoretische Physik, ETH Z\"{u}rich, 8093 Z\"{u}rich, Switzerland}
\author{Thomas Maier}
\affiliation{Center for Nanophase Materials Sciences and Computer Science and Mathematics Division, Oak Ridge National Laboratory, Oak Ridge, TN 37831-6494}
\date{\today }
\begin{abstract}
We present a  submatrix update 
algorithm for the continuous-time auxiliary field method that allows the simulation of large lattice and impurity problems. The 
algorithm
takes optimal advantage of modern CPU architectures by consistently using 
matrix instead of vector 
operations, resulting in a speedup of a factor of $\approx 8$ and thereby 
allowing access to larger systems and lower temperature. 
We illustrate the power of our algorithm at the example
of a cluster dynamical mean field simulation of the N\'{e}el transition in the 
three-dimensional Hubbard model, where we show momentum dependent self-energies for clusters with up 
to 100 sites.
\end{abstract}


\maketitle 
The theoretical investigation of correlated fermionic lattice
systems has been one of the most challenging tasks in condensed matter physics.
Many of these systems are not tractable with controlled analytic approximations
in the regimes of interest, so that numerical simulations need to be employed.
Several numerical approaches exist: With exact diagonalization\cite{Dagotto91} 
(ED) one calculates the exact eigenstates of a system on a small lattice.
Because the Hilbert space grows exponentially with lattice size, ED is
limited to comparatively small systems.  Variational methods like the density
matrix renormalization group theory\cite{White92,Schollwoeck05} (DMRG) work
well in one dimension, but extensions to two-dimensional systems
\cite{PEPS,Vidal05,Schuch08,Corboz10} are still under development.  Standard
lattice Monte Carlo methods\cite{BSS81} are hampered by the fermionic sign
problem\cite{Loh90,Troyer05} that limits access to large system size or low
temperature away from half filling.

Systems with a large coordination number are often studied within the dynamical
mean field approximation (DMFT)\cite{Georges96,Kotliar06}.  Early studies by
Metzner and Vollhardt\cite{Metzner89} and M\"{u}ller-Hartmann\cite{MH89} showed
that the diagrammatics of interacting fermions becomes purely local in the 
limit of infinite coordination number. In this case the solution of the
lattice model may be obtained from the solution of an impurity model and an
appropriately chosen self-consistency condition\cite{Georges92b}.

Later work on cluster extensions of DMFT
\cite{Hettler98,Lichtenstein00,Hettler00,Kotliar01,Maier05} took into account
non-local correlations in addition to the local correlations already contained 
within the DMFT by considering ``cluster'' impurity models with an internal
momentum structure \cite{Maier05}. These cluster approximations are based on
a self-energy expansion in momentum space, $\Sigma(k, \omega) \approx
\sum_K^{N_c} \Sigma_K(\omega) \phi_K(k)$\cite{Okamoto03} that becomes exact in
the limit of a complete momentum space basis ($N_c \rightarrow \infty$) and can
therefore be controlled by increasing the cluster size.

Quantum impurity models are well suited to numerical study. Methods for their
solution include numerical renormalization group approaches\cite{Bulla08},
exact diagonalization\cite{Caffarel94}, and approximate semi-analytical
resummation of classes of diagrams\cite{Georges92b,Pruschke89,Coleman84}.
However, until a few years ago only the Hirsch-Fye quantum Monte
Carlo\cite{Hirsch86} algorithm was able to obtain unbiased and numerically
exact solutions of large cluster impurity problems at intermediate interaction
strength.  This changed with the development of continuous-time methods
\cite{Rubtsov04,Rubtsov05,Werner06,Werner06Kondo,Gull08_ctaux,laeuchli09}.  The
vastly better scaling\cite{Gull07} of these methods and the absence of
discretization errors allowed access to lower temperatures, larger
interactions, and more orbitals.

Large cluster calculations remain computationally challenging as the numerical
cost -- even in the absence of a sign problem -- scales as $O[(N_c\beta U)^3]$
in the case of the interaction expansion\cite{Rubtsov05,Gull08_ctaux}, and
$O[\exp(N_c) \beta^3]$ in the hybridization expansion\cite{Werner06Kondo}
methods (for single orbital cluster Anderson models at inverse temperature
$\beta$ and interaction $U$ for a cluster of size $N_c$). It is therefore
important to develop efficient algorithms to solve cluster impurity models.

Two numerical algorithmic improvements have significantly increased the size of
systems accessible by simulations with the Hirsch-Fye algorithm: the
``delayed'' updates\cite{Alvarez08}, and the ``submatrix''
updates\cite{Nukala09}. An important question is therefore if these techniques
may be generalized to the continuous-time algorithms and whether similar
savings in computer time may be expected, and how these savings translate into
newly accessible physics.

Both ``delayed'' and ``submatrix'' updates are  mainly based on efficient memory
management; ``submatrix'' updates
further reduce the algorithmic complexity of the updating
procedure.
Modern computer architectures employ a memory hierarchy:
Calculations are performed on data loaded into registers. Any data that are not
in the registers are stored either in the ``cache'' (currently with a size of a few
MB) or in the ``main memory'' (with a size of a few GB). The cache is relatively fast,
but there is little of it, while access to the main memory is often slow and shared
among several compute cores. The bottleneck in many modern scientific
applications, including the continuous-time algorithms, is not the speed at
which computations are performed, but the speed at which data can be loaded from
and stored into main memory. 

\begin{figure}[htb]
\begin{flushleft}
\subfigure[]{\label{shermansketch}\includegraphics[height=2.1cm]{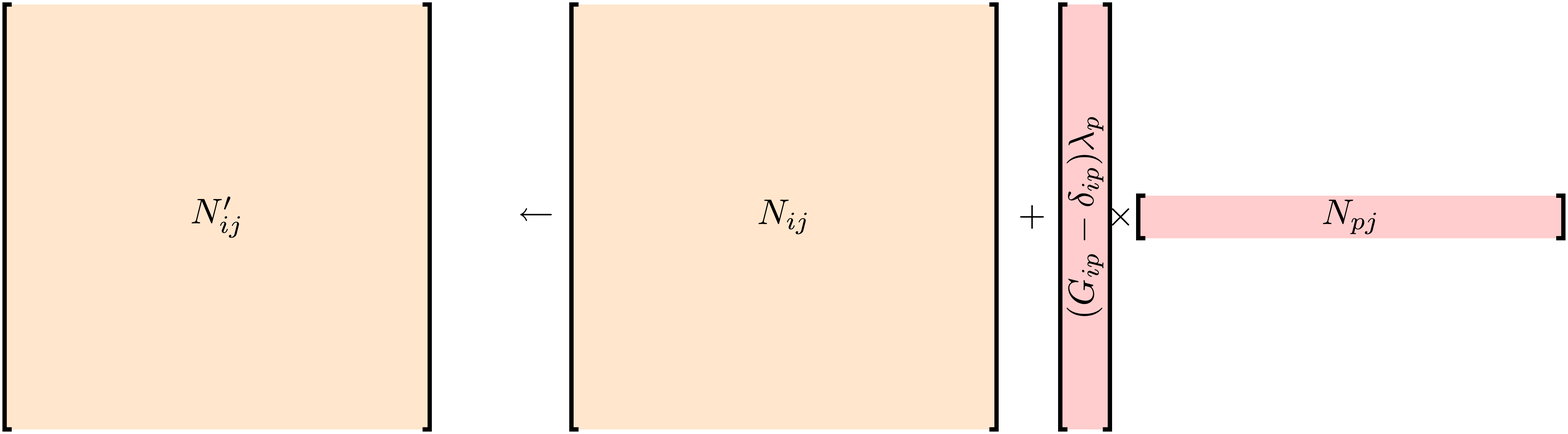}}
\subfigure[]{\label{woodburysketch}\includegraphics[height=2.1cm]{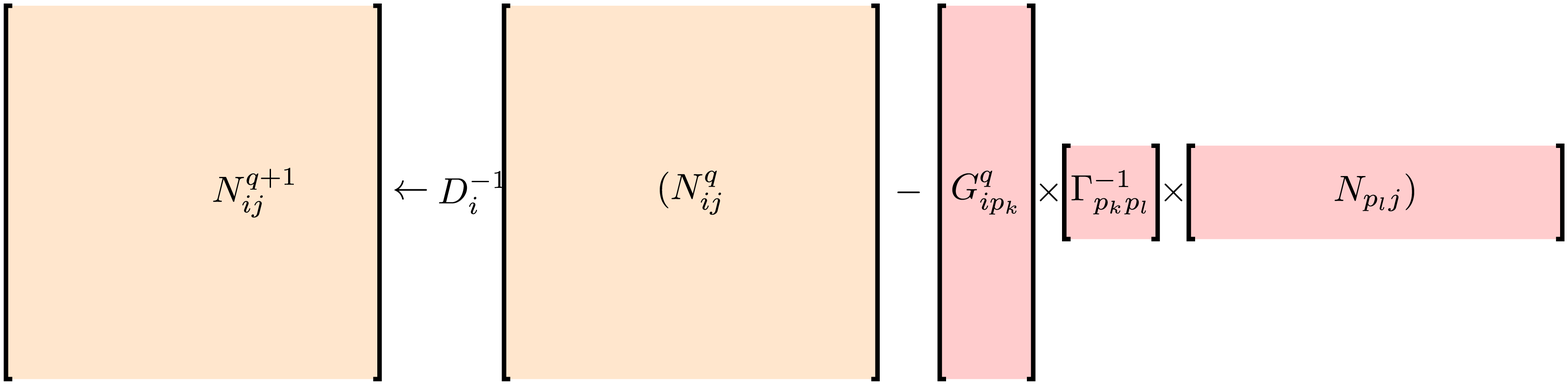}}
\end{flushleft}
\caption{(Color online) Illustration of update formulas. \ref{shermansketch}: ``rank-one'' 
updates of Ref.~\onlinecite{Gull08_ctaux}, accessing $O(m^2)$ data points for 
$O(m^2)$ operations and performing one update. \ref{woodburysketch}: submatrix 
updates, accessing $O(m^2)$ values but performing $O(m^2k)$ operations, for $k$ 
updates.
}
\label{sketch}
\end{figure}

The central object in continuous-time algorithms is a matrix, which for large
cluster calculations does not fit into the cache.  Monte Carlo updates often
consist of rank-one updates or matrix-vector products.  Such updates perform
$O(m^2)$ operations on $O(m^2)$ data, where $m$ is the average matrix size, and
therefore run at the speed of memory. Matrix-matrix operations [with $O(m^3)$
operations executed on $O(m^2)$ data] could run at the speed of the registers,
as more (fast) calculation per (slow) load / store operation are performed.
The reason behind the success of both the ``submatrix'' and the ``delayed''
updates is the combination of several (slow) successive rank-one operations
into one fast matrix-matrix operation, at the cost of some minimal overhead.
This is illustrated in Fig.~\ref{sketch}.

The delayed update algorithm can be straightforwardly generalized to
(non-ergodic) spin-flip operations in the interaction expansion (CT-INT) and
continuous-time auxiliary field algorithms (CT-AUX)\cite{Werner09_8site}, and
an adaptation of the concept of delayed updates to vertex insertion and
removals in the interaction expansion was recently proposed by Mikelsons\cite{KarlisThesis}.

In this article we present a generalization of the
``submatrix'' technique of Ref.~\onlinecite{Nukala09} to the CT-AUX algorithm,
which uses fewer redundant operations than ``delayed'' updates.
We find a speed increase of $\approx$ $8$ for a typical large cluster impurity problem. 
We demonstrate the scaling both as a function of computational resources and as a function of problem size, and we show results for controlled large-scale cluster calculations.

The paper is structured as follows: In Sec.~\ref{auxsec} we reintroduce the
CT-AUX algorithm and describe the Monte Carlo random walk procedure. In
Sec.~\ref{subsec} we introduce the submatrix updates, and in Sec.~\ref{impsec}
we apply them to CT-AUX. Section \ref{ressec} shows physics and benchmarking
results for the new algorithm, and Sec.~\ref{concsec} contains the conclusions.

\section{The Continuous-Time Auxiliary Field algorithm}\label{auxsec}
We present the submatrix updates for
CT-AUX\cite{Gull08_ctaux}, for which the linear algebra is similar to the
well-known Hirsch Fye\cite{Hirsch86} method. 
To introduce notation and
conventions we repeat the important parts of the derivation of
Ref.~\onlinecite{Gull08_ctaux}, limiting ourselves to the description of the
dynamical mean field solution of the single orbital Anderson impurity model.
{\it Lattice} problems [i.e., problems without hybridization terms in the Hamiltonian and with no (cluster) dynamical mean field self-consistency imposed] differ only in the form of the non-interacting Green's function. Their simulation proceeds along the same lines and will not be treated separately here.

\subsection{Partition Function Expansion}
The Hamiltonian of the single orbital Anderson impurity model describes the behavior of an impurity (described by operators $d_\sigma, d_\sigma^\dagger$) with an on-site energy $\epsilon_0$ and on-site interaction $U$ coupled by a hybridization with strength $V_{p\sigma}$ to a bath (described by $a_{p\sigma},a_{p\sigma}^\dagger$) with dispersion $\epsilon_p$:
\begin{align}
H &=H_{0} + V, \label{Ham}\\
H_0 &=-(\epsilon_0-U/2)(n_\uparrow+n_\downarrow)\nonumber\\
&+\sum_{\sigma, p} (V_{p\sigma} d^\dagger_\sigma a_{p\sigma} + H. c.) + \sum_{\sigma,p} \epsilon_p a^\dagger_{p\sigma}a_{p\sigma},\\
V &= U  \left[n_\uparrow n_\downarrow - \frac{n_\uparrow +n_\downarrow}{2}\right].
\end{align}
$n_\sigma=d_\sigma^\dagger d_\sigma$ denotes the impurity occupation.
Continuous-time algorithms expand expressions for the partition function $Z=\Tr \exp(-\beta H)$ (at inverse temperature $\beta$) into a diagrammatic series. In CT-AUX, the  series
is a perturbation expansion in the interaction:
\begin{align}
Z &= \sum_{n\geq 0}
\int_0^\beta d\tau_1 \ldots  \int_{\tau_{n-1}}^\beta \!\!\!\! d\tau_n 
\Big( \frac{K}{\beta}\Big)^n  
\Tr \Big[ e^{ -(\beta-\tau_{n}) H_{0}} \nonumber\\
&\times\Big( 1- \frac{\beta V}{K}  \Big) 
\ldots
e^{ - (\tau_{2} -\tau_{1}) H_{0}} \Big( 1- \frac{\beta V}{K} \Big) 
e^{ -\tau_{1} H_{0}}
\Big].
\end{align}
The interaction term $V$ in this expansion can be decoupled with an auxiliary 
field\cite{Rombouts99}
\begin{subequations}
\begin{align}
1-\frac{\beta V}{K}&= \frac{1}{2}\sum_{s=-1,1}e^{\gamma s (n_\uparrow-n_\downarrow)},\label{expansion_parameter}\\
\cosh(\gamma)&\equiv 1+(\beta U)/(2K),\label{gamma}
\end{align}
\end{subequations}
introducing an arbitrary constant $K$ and auxiliary ``spins'' $s$. Hence
\begin{align} \label{decoupledZ}
Z= \sum_{n\geq 0}
\sum_{s_i= \pm 1 \atop 1\leq i\leq n}
\int_0^\beta d\tau_1 \ldots  \int_{\tau_{n-1}}^\beta \!\!\!\! d\tau_n
\Big( \frac{K}{2\beta}\Big)^n    Z_{n}, \end{align}\begin{align}
Z_{n}(\stau) \equiv \Tr 
\prod_{i=n}^{1} %\prod_{i=1}^{n}
e^{ - \Delta \tau_{i} H_{0} } 
e^{s_{i} \gamma (n_{\uparrow} - n_{\downarrow})  }.\label{Zn}
\end{align}
Note that the insertion of an arbitrary number of ``interaction vertices'' 
(auxiliary spin and time pairs) $(s_j, \tau_j)$ with $s_j=0$ 
into Eq.~(\ref{decoupledZ}) does not change the value of $Z_{n}(\stau).$ We will refer to auxiliary spins with value $s_n=0$ as ``non-interacting'' spins.

We can express the trace of exponentials of one-body operators in 
Eq.~(\ref{decoupledZ}) as a determinant of a $(n\times n)$ matrix $N$,
\begin{align}
\frac{Z_{n}(\stau) }{Z_0} &=& \prod_{\sigma=\uparrow,\downarrow} \det 
N_\sigma^{-1}(\stau),
\label{Zn_div_Z0}\\
N^{-1}_{\sigma}(\stau) &\equiv& e^{V_{\sigma}^{\{s_{i}\}}} 
-\mathcal{G}_{0\sigma}^{ \{\tau_{i}\}} \label{Nmatrix}
\Big(e^{V_{\sigma}^{ \{s_{i}\} }} - 1 \Big),
\\
e^{V_{\sigma}^{\{s_{i}\}}}&\equiv&\text{diag}\Big(e^{\gamma (-1)^{\sigma }s_1}, \ldots, e^{\gamma (-1)^{\sigma } s_n}\Big).
\end{align}
$G_{0\sigma}^{\{\tau_i\}}$ denotes a $(n\times n)$ matrix of bare Green's 
functions, 
$(\mathcal{G}_{0\sigma}^{\{\tau_i\}})_{ij}=\mathcal{G}_{0\sigma}(\tau_i-\tau_j)$. From now on we will omit the spin index $\sigma$.

The matrix $N$ is related to the Green's function matrix $G$ by $G = N G_0$. 
The matrices $G$ and $N$ for auxiliary spin configurations that have the same imaginary time 
location for all vertices, but differ in the value of an auxiliary spin $s_p$, are related by a Dyson equation
\begin{subequations}
\begin{align}
N_{ij}' &= N_{ij} + (G_{ip} - \delta_{ip}) \lambda N_{pj},\label{DysonG}\\
G_{ij}' &= G_{ij} + (G_{ip} - \delta_{ip}) \lambda G_{pj},\label{DysonN}\\
\lambda &= e^{V_p'-V_p} -1.
\end{align}
\end{subequations}
This relation is the basis for spin-flip updates.

\subsection{Random Walk} The infinite sum over expansion orders $n$ and the
integral and sum over vertices $\{(s_i,\tau_i)\}$ in Eq.~(\ref{decoupledZ}) is
computed to all orders in a stochastic Monte Carlo process: The algorithm
samples time ordered configurations $\{(s_i,\tau_i)\}$ with weight
\begin{equation} 
w(\stau)=\Big(\frac{K d\tau}{2\beta}\Big)^n
\prod_{\sigma=\uparrow,\downarrow} \det N_\sigma^{-1}(\stau).  
\end{equation}

To guarantee ergodicity of the sampling it is sufficient to insert and remove spins with a random orientation $s_i=\uparrow,\downarrow$ at random times $0 \leq \tau_i <\beta$. 
Spin insertion updates are balanced by removal updates. For an insertion update we select a random time in the interval $[0,\beta)$ and a random direction for this new spin, leading to a proposal probability $p^\text{prop}(n\rightarrow n+1)=(1/2)(d\tau/\beta$). For removal updates a random spin is selected and proposed to be removed, leading to a proposal probability $p^\text{prop}(n+1\rightarrow n)=1/(n+1)$.
The combination of Eq.~(\ref{Zn_div_Z0}) with these proposal probabilities leads to the Metropolis acceptance rate $p(n\rightarrow n+1)$ $=$ $\min(1, R)$ with
\begin{align}\label{insprob}
R=\frac{K}{n+1}\prod_{\sigma=\uparrow,\downarrow} \frac{\det [N^{(n+1)}_\sigma]^{-1}}{\det [N^{(n)}_\sigma]^{-1}},
\end{align}
where $(n)$ denotes the dimension of $N_\sigma^{-1}$.

In addition to the insertion and removal updates we consider spin flips of auxiliary spins. These updates are self-balancing, and the transition probability from a state $\stauf$ to a state $\stauprimef$ is given by
\begin{align}
R=\prod_{\sigma=\uparrow,\downarrow} \frac{\det [N^{(n)}_\sigma(\stauprimef)]^{-1}}{\det [N^{(n)}_\sigma(\stauf)]^{-1}}.
\end{align}

In the particle hole symmetric case the parameter $K$ may be chosen such that only even orders in the perturbation series occur and that the average perturbation order is half as large as the one of the algorithm presented here (see Ref.~\onlinecite{Werner10} for details in the real-time context, where this scheme allowed propagation to much longer times). As the resulting algorithm is less general and requires double-vertex insertions it will not be explored here.

Non-interacting auxiliary spins, or auxiliary spins with value $0$, do not change the value of $Z_n$ in Eq.~\ref{Zn}. 
We will make use of this fact to precompute a matrix that is equivalent to $N$ but contains non-interacting vertices represented by spin $0$ auxiliary spins. Insertion and removal updates then become equivalent to spin-flip updates (from $0$ to $1$ or $-1$ and vice versa), thus allowing for a similar application of the sub-matrix update algorithm as in the case of the Hirsch-Fye solver \cite{Nukala09}. This procedure is explained in more detail in Sec.~\ref{impsec}.

\section{Submatrix Updates}\label{subsec}
To derive the sub-matrix updates\cite{Nukala09} let us consider a typical step $k$ of the algorithm at which the interaction $p_k$ [with spin and time ($s_{p_k}$, $\tau_{p_k}$) of $m$ interaction vertices] is changed from $V_{p_k}$ to $V_{p_k}'$.
The new matrix $G^{k+1}$ is then given by Eq.~(\ref{DysonG}),
\begin{align}
G_{ij}^{k+1} &= G_{ij}^k + (G_{ip_k}^k - \delta_{ip_k}) \lambda^k G_{p_kj}^k,\\
\lambda^k &= e^{V_{p_k}'-V_{p_k}}-1.\nonumber
\end{align}
$\lambda^k$ denotes the change of interaction at step $k$.
We proceed by showing how the determinant ratio $\det N^k/\det N^{k+1}$ of Eq.~(\ref{insprob}) as well as the new matrix $N^{k+1}$ are
computed efficiently using the Woodbury formula: We define an inverse matrix $A$ of $G$, analyze its changes during an update, and show 
how they can be incorporated in a small ($k\times k$) matrix $\Gamma$ that is easily computed by accessing only $k^2 \ll m^2$ matrix elements in each step. 
The inverse of this matrix is then iteratively computed either by employing an $LU$ decomposition%, as in the paper on submatrix updates for the Hirsch Fye algorithm
, or a partitioning scheme.

A change to the inverse Green's function matrix $A^k = (G^k)^{-1}$ is of the form\cite{ShermanMorrison50}
\begin{align}\label{Ainv}
A^{k+1}_{ij} &= A^k_{ij} + \gamma^k (A^k_{ip}-\delta_{ip})\delta_{pj} \\
&=A^{k}_{ij} + \gamma^k A^k_{ip}\delta_{pj} - \gamma^k\delta_{ip}\delta_{pj},\nonumber\\ \nonumber
\gamma^k &= e^{-\gamma \sigma (s_{p_k}'-s_{p_k})}-1.
\end{align}
$\gamma^k$, similar to $\lambda^k$ above, contains the information about the changed interaction at step $k$.
Eq.~(\ref{Ainv}) is commonly known as the Sherman Morrison formula and illustrated in Fig.~\ref{shermansketch}.
We define $\tilde{A}^k_{ij} = A^k_{ij} + \gamma^k A^k_{ip}\delta_{pj}$, i.e. the matrix $A^k$ where the $p$-th column is multiplied by $(1+\gamma^k)$, and therefore $\det \tilde{A}^k = (1+\gamma^k)\det(A^k)$.
We then rewrite Eq.~(\ref{Ainv}) as $A^{k+1}_{ij} = \tilde{A}^k_{ij}-\gamma^k\delta_{ip}\delta_{pj},$ %$\tilde{G} = \tilde{A}^{-1},$
and, using the ``matrix determinant lemma'' $\det(A_{ij}+u_i v_j) = [1 + v_l (A^{-1})_{lq} u_q]\det A_{ij}$, we have 
\begin{align}\label{Akp1Ak}
\det A^{k+1} &= \det(\tilde{A}^k)\det (1-\gamma^k [(\tilde{A}^k)^{-1}]_{pp})\\
&=\det A^k \left(1+\gamma^k\right)\left(1-\frac{\gamma^k}{1+\gamma^k}G^k_{pp}\right)\nonumber \\\nonumber
&=-\det A^k \gamma^k \left[G^k_{pp}-\frac{1+\gamma^k}{\gamma^k}\right].
\end{align}
This formula yields the determinant ratio
\begin{align}
\frac{\det N^k}{\det N^{k+1}} = -\gamma^k \left[G^k_{pp}-\frac{1+\gamma^k}{\gamma^k}\right]
\end{align}
needed in Eq.~(\ref{insprob}) for the acceptance or rejection of an update.

We can recursively apply Eq.~(\ref{Akp1Ak}) to obtain an expression for performing multiple interaction changes, as long as they occur for different spins $p_i \neq p_j (i\neq j)$:
\begin{align}
A^{k+1}_{ij} &= A^0_{ij} + \sum_{l=0}^k \gamma^l (A^0_{ip_l} -\delta_{ip_l})\delta_{p_l j}\nonumber \\ &=\tilde{A}^k_{ij} - \sum_{l=0}^k \gamma^l \delta_{ip_l}\delta_{p_lj} \nonumber\\
&= \tilde{A}^k - X^k (Y^k)^T, \label{Aknew}\\
X^k_{ij}&=\gamma_j\delta_{ip_j},\\
(Y^k)^T_{ij}&=\delta_{p_ij}.
\end{align}
The new matrix $A^{k+1}$ is therefore generated from $A^0$  by successively multiplying columns $p_l, 0 \leq l \leq k$ of $A^0$ with $\gamma^l$ 
and adding constants to the diagonal. $X$ and $Y^T$ are index matrices that label the changed spins and keep track of a prefactor $\gamma^k$.

For measurements we need access to the Green's function $G$, not its inverse $A$. It is obtained after $k_\text{max}$ 
steps by applying the Woodbury formula Eq.~(\ref{Woodbury}) to Eq.~(\ref{Aknew}): with $q$ denoting a Woodbury step combining $k_\text{max}$ vertex update steps:
\begin{align}
G^{q+1} &= (A^{q+1})^{-1} \nonumber \\ &= \tilde{A}^{-1} + \tilde{A}^{-1} X(1-Y^T \tilde{A}^{-1} X)^{-1}Y^T \tilde{A}^{-1}, \label{Woodbury}\\
G^{q+1} &= \tilde{G}+ \tilde{G} X(1-Y^T \tilde{G} X)^{-1}Y^T \tilde{G}, \label{Gq}
\end{align}
where $\tilde{G} = \tilde{A}^{-1}$. After some simplification, Eq.~(\ref{Gq}) can be shown to be 
\begin{align}
G^{q+1}_{ij} &= D^{-1}_i \left(G_{ij} - G_{ip_k} \Gamma^{-1}_{p_kp_l} G_{p_lj}\right). \label{Gshermanmorrison}
\end{align}
Here we have introduced a $k_\text{max}\times k_\text{max}$ - matrix $\Gamma,$ defined as 
\begin{align}
\Gamma_{pq} = G^0(p,q) -\delta_{pq}\frac{1+\gamma_p}{\gamma_p},\label{Gamma}
\end{align}
and a vector $D$ that is $1$ everywhere but at positions where auxiliary spins are changed:\begin{align}
D_{p_k}^{-1} &= \frac{1}{1+\gamma^k}.
\end{align}
Note that $G^0$ is the interacting Green's function at step $k=0$ and not the bare Green's function $\mathcal{G}^0$ of the effective action, unless all auxiliary spins are zero.

Translating this Green's function formalism to a formalism for $N$ matrices is straightforward: writing $G=N\mathcal{G}^0$ and
multiplying Eq.~(\ref{Gshermanmorrison}) from the right with $(\mathcal{G}^0)^{-1}$ yields 
\begin{align}
N^{q+1}_{ij} &= D^{-1}_i (N_{ij} - G_{ip_k} \Gamma^{-1}_{p_kp_l} N_{p_lj}),\label{Nshermanmorrison}
\end{align}
where one $G_{ip_k}$ remains in Eq.~(\ref{Nshermanmorrison}). This equation is illustrated in Fig.~\ref{woodburysketch}.

Inserting $G = N G_0$ into Eq.~(\ref{DysonG}) and setting $V'=0$ $(N' = 1)$ we obtain:
\begin{align}
1 &= N e^V - N G_0 e^V + N G_0 \\
(N G_0)_{ij} &= (N_{ij} e^{V_j} -\delta_{ij})/(e^{V_j}-1) =  G_{ij}\\
N_{ij} &= G_{ij}(1-e^{-V_j})+e^{-V_j}\delta_{ij}. \label{NfromG}
\end{align}
The computation of $G$ from $N$ in this manner fails if the interaction $V_j$ is zero. In this case we need to compute $G_{ij} = N_{ik} \mathcal{G}^0_{kj}$ at a cost of $O(N)$ for each $i$ and $j$.

\subsection{Determinant Ratios and Inverse Matrices}
To either accept or reject a configuration change, we need to compute the determinant ratio $\det N^{k+1}/\det N^k$ [Eq.~(\ref{insprob})]. Following Ref.~\onlinecite{Nukala09} we write:
\begin{align}
\det A^{k+1} = (-1)^{k+1} \prod_{j=0}^k \gamma_j \det A^0 \det \Gamma^k.
\end{align}
The computation of the determinant $\det \Gamma^k$ is an expensive $O(k^3)$ operation, if $\Gamma^k$ has to be recomputed from scratch. However,
we successively build $\Gamma^k$ by adding rows and columns. In the following we present two efficient (and as far as we could see equivalent) methods to iteratively compute determinant ratios of $\Gamma$: keeping track of an $LU$ decomposition, and storing the inverse computed using inversion by partitioning.

\subsubsection{$LU$ decomposition}
For each accepted update we keep track of a $LU$ decomposition of $\Gamma$:
\begin{align}
\Gamma^k &= \begin{pmatrix} \Gamma^{k-1} &s \\ w^T &d \end{pmatrix} = \begin{pmatrix} L^{k-1}&0 \\ x^T&1\end{pmatrix} \begin{pmatrix} U^{k-1}&y \\ 0 &\beta\end{pmatrix},\\
L y &=s,\label{lineq1}\\ 
U^T x &=w,\label{lineq2}\\
\beta &= G^0(p^k,p^k) - \frac{1+\gamma^k}{\gamma^k}-x^Ty\label{detrat}
\end{align}
where both $x^T$ and $y$ are computed in $O(k^2)$ by solving a linear equation for a triangular matrix.
The determinant ratio needed for the acceptance of an update is 
\begin{align}
\frac{\det A^{k+1}}{\det A^k} = -\beta \gamma^k.\label{betagammak}
\end{align}

These updates have been formulated for spins that have only been updated once. In the case where the same spin is 
changed twice or more, rows and columns in $\Gamma$, or $L$ and $U$, need to be modified. These changes are of the form $\Gamma \rightarrow \Gamma + uv^T$, 
and Bennett's algorithm \cite{Bennett65} can be used to re-factorize the matrix. 

The probability to accept/reject a $(k+1)$-th spin requires $O(k^2)$ operations [computation of 
$x$ and $y$ using Eqs.~(\ref{lineq1}) and (\ref{lineq2}) requires $O(k^2)$ operations, while 
Eq.~(\ref{detrat}) requires $O(k)$ operations]. On the other hand, the ``delayed" algorithm 
requires $O(km)$ operations to compute the acceptance rate of a $(k+1)$-th spin flip, for a matrix of size $m$.
In this sense, the submatrix update methodology not only manages matrix operations efficiently, 
but also improves the computational efficiency of the spin-flip acceptance rate. 

\subsubsection{Inversion by Partitioning}
Alternatively, we can compute the inverse of $\Gamma$ by employing the Sherman-Morrison formula:
\begin{align}
\beta &= (d - w^T\Gamma_k^{-1}s)\\
\Gamma_{k+1}^{-1} &= \begin{pmatrix} \Gamma_k^{-1} + (\Gamma_k^{-1} s){\beta^{-1}}(w^T\Gamma_k^{-1}) & -\Gamma_k^{-1} s\beta^{-1} \\
-\beta^{-1}w^T\Gamma_k^{-1} & \beta^{-1}\end{pmatrix},%\\
\end{align}
\begin{align}
\frac{\det\Gamma^{k+1}}{\det\Gamma^k}&=\beta,\qquad\frac{\det A^{k+1}}{\det A^k} = -\gamma^k\beta.\label{ratS}
\end{align}
Although both methods obtain the acceptance rates of Eqs.~(\ref{betagammak}) and (\ref{ratS}) in $O(k^2)$ steps, 
inversion by partitioning requires an additional step of updating the $\Gamma_{k+1}^{-1}$, and hence is 
expected to be slower than the $LU$ decomposition approach. However, the complication of re-orthogonalizing the $LU$ factorized matrix using Bennett's algorithm does not arise.

\section{The random walk with submatrix updates}\label{impsec}
The sums and
integrals of Eq.~(\ref{decoupledZ}) are computed by a random walk in the space
of all expansion orders, auxiliary spins, and time indices. In the cluster
case, configurations acquire an additional site index. A configuration $c_k$ at
expansion order $n$ contains $n$ interaction vertices with spins, sites, and
time indices:
\begin{align}
c_k = \{(\tau_1, s_1, \sigma_1), \cdots (\tau_n, s_n, \sigma_n)\}.
\end{align}
The configuration space $\mathcal{C}$ consists of all integrands / summands in
Eq.~(\ref{decoupledZ}), which we can represent by sets of triplets of numbers,
consisting of auxiliary spins, times, and site indices:
\begin{align}
\mathcal{C} = \{c_0, \cdots, c_k((\tau_1, s_1, \sigma_1), \cdots, (\tau_k, s_k, \sigma_k)), \cdots\}.
\end{align}
%We employ a random walk in this configuration space to sample the integral
%Eq.~(\ref{decoupledZ}):  starting from an initial configuration we propose to
%insert, remove, or change vertices and we thereby sample the integral
%stochastically using a Metropolis Monte Carlo procedure. We propose the
%following updates for the random walk: the insertion of a vertex at a random
%site and time with a random auxiliary spin, the removal of a previously
%existing vertex, and the spinflip of an auxiliary spin. The first two updates
%are required to sample the configuration space, and the last (spinflip)
%updates are mentioned here because they can be computed relatively cheaply.

To efficiently make use of the submatrix updates, we add an additional
step before insertion and removal updates are performed. In this
preparation step, we insert a number $k_\text{max}$ of randomly chosen
non-interacting vertices with auxiliary spin $s=0$, which, as
discussed in Sec.~\ref{auxsec}, does not change the value of the
partition function. Once these vertices are inserted, insertion and
removal updates at the locations of the pre-inserted non-interacting
vertices become identical to spin-flip updates: an insertion
update of a spin $s=1$ now corresponds to a spin-flip update from spin
$s=0$ to spin $s=1$, and similar for removal updates. This
pre-insertion step of non-interacting vertices then allows for a
similar application of submatrix updates as in the case of the
Hirsch-Fye algorithm. 

To accommodate this pre-insertion step, we split our random walk into an inner
and an outer loop. In the outer loop (labeled by $q$) we perform measurements
of observables and run the preparation step discussed above as well as
recompute steps. These steps are described in more detail below. In the inner
loop (labeled by $k$) we perform $k_\text{max}$ insertion, removal, or spin-
flip updates at the locations of the pre-inserted non-interacting spins.  It is
best to choose $\langle m\rangle \gg k_\text{max} \gg 1$ so the blocking
becomes efficient, but matrices of linear size $k_\text{max}$ are small enough to fit
into the cache. 

\subsection{Preparation steps} We begin a Monte Carlo sweep
with preliminary computations for spins that we will propose to insert or remove. For
this, we generate randomly a set of $k^\text{ins}_\text{max}$ pairs of (site,
time) indices, where $k^\text{ins}_\text{max}$ denotes the maximum insertions
possible. We then compute the additional rows of the matrix $N$ for these
noninteracting spins:
\begin{align} \tilde{N} = \begin{pmatrix} N &0 \\\tilde{R}  & 1\end{pmatrix},
\end{align}
where $\tilde{R}$ is a matrix of size $n \times k^\text{ins}_\text{max}$
containing the contributions of newly added noninteracting spins,
\begin{align}
\tilde{R}_{ij} = \mathcal{G}^0_{ik}(e^{-\gamma \sigma s_k} -1)N_{kj},
\end{align}
at the cost of $O(n^2k^\text{max}_\text{ins}),$ as well as the Green's function
matrix $G=N\mathcal{G}^0$ for the new spins (cost
$n^2k^\text{max}_\text{ins}$).

\subsection{Insertion, removal, spinflip of an auxiliary spins}
Vertex insertion updates are performed by proposing to flip one of the newly inserted non-interacting spins from 
value zero to either plus or minus one. The determinant ratio is obtained by using Eqs.~(\ref{lineq1}), (\ref{lineq2}), (\ref{betagammak}),
and (\ref{detrat}), (i.e., by the solution of a linear equation of a triangular matrix). If the update is accepted
the auxiliary spin is changed and the matrix $\Gamma$ is enlarged by a row and a column.

Starting from a configuration $c_k = \{(\tau_1, s_1, \sigma_1), \cdots (\tau_k, s_k, \sigma_k)\}$ we propose to remove the interaction vertex $ (\tau_j, s_j, \sigma_j)$. 
The ratio of the two determinants [Eq.~(\ref{detrat})] is computed by proposing to flip an auxiliary spin from $\pm 1$ to zero. 
For this we compute $s$ and $w$ as in Eq.~(\ref{Gamma}), and then compute $x$ and $y$ by solving a linear equation for a triangular system [Eqs.~(\ref{lineq1}) and (\ref{lineq2})].
Finally, Eq.~(\ref{betagammak}) is computed using Eq.~(\ref{detrat}).
If the update is accepted the auxiliary spin is set to zero and  $\Gamma$ is enlarged by a row and a column.

Double vertex updates required for the scheme of Ref.~\onlinecite{Werner10} proceed along the same lines and enlarge $\Gamma$ by two rows and two columns.

To perform a spin-flip update we choose a currently interacting spin with value $\pm 1$ and propose to flip it to $\mp 1$ using
Eqs.~(\ref{lineq1}), (\ref{lineq2}), and (\ref{betagammak}). If the update is accepted, $\Gamma$ grows by a row and a column.

\subsection{Recompute step}
This scheme of insertion, removal, and spinflip updates is repeated $k_\text{max}$ times. With each accepted move the matrix $\Gamma$ grows by a row and a column.% To keep the numerics efficient we recompute the full $N$ after $k_\text{max}$ steps,
To keep the algorithm efficient we periodically recompute the full $N$ matrix
using the Woodbury formula \ref{Nshermanmorrison}:
\begin{align}
N^{q+1}_{ij} &= D^{-1}_i (N_{ij} - G_{ip_k} \Gamma^{-1}_{p_kp_l} N_{p_lj}),\label{Gshermanmorrison2}
\end{align}
as $\Gamma$ grows with every accepted update, and the cost of computing determinant ratios is $O(k^2)$.
The recompute step consists of two inversions for $L$ and $U$, which are both $O(k^2)$ operations, and two matrix multiplications, at cost $O(k^2N)$ and $O(N^2k)$ respectively.
Noninteracting auxiliary spins can then be removed from $N^{q+1}_{ij}$ by deleting the corresponding rows and columns.

\subsection{Measurements}\label{meassec}
At the end of a sweep, if the system is thermalized, observable averages are computed. As the complete $N$-matrix is known at this point, the formulas presented in Ref.~\onlinecite{Gull08_ctaux} are employed without change. In most calculations, the computation of the Green's function is the most expensive part of the measurement. In large ``dynamical cluster approximation'' (DCA)\cite{Hettler98,Hettler00,Maier05} calculations it is therefore advantageous to compute directly the Green's functions in cluster momenta, of which there are only $N_c$, in contrast to the $N_c^2$ real-space Green's functions. Also, on large clusters, Green's functions are best measured directly in Matsubara frequencies.

\section{Results}\label{ressec}
We present two types of results. First we examine the performance of submatrix updates in practice, using several scaling metrics. We then illustrate a physics application where we test the DCA
approximation on large clusters, showing cluster size dependence and extrapolations to the infinite system limit.

\begin{figure}[tb]
\includegraphics[width=0.8\columnwidth]{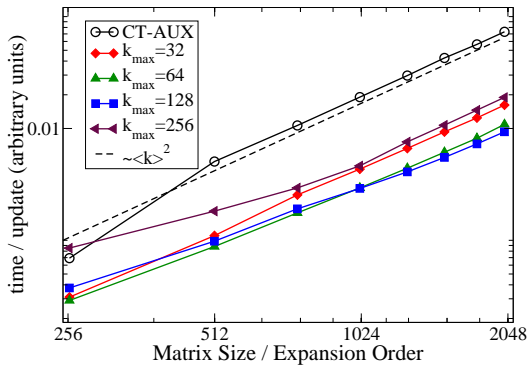}
\caption{Time per update (in arbitrary units) for submatrix and rank one CT-AUX updates. Open circles (black online): rank one updates. Filled diamonds, triangles, squares, and left triangles: submatrix updates for $k_\text{max}=32, 64, 128,$ and $256$. Dashed line: ideal $O(k^2)$ scaling, arbitrary prefactor.}
\label{SecUpdate}
\end{figure}
\begin{figure}[tb]
\includegraphics[width=0.8\columnwidth]{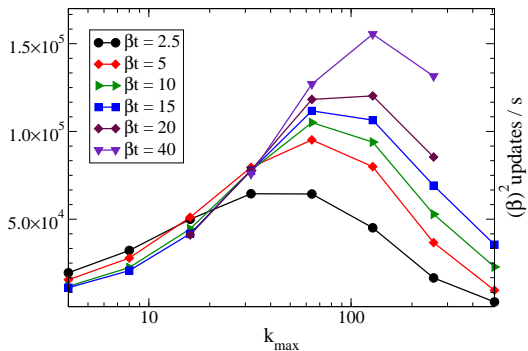}
\caption{Updates per time as a function of $k_\text{max}.$ $16$-site cluster, $U/t = 8$, for temperatures indicated.}%, \beta t=20, t'/t = -0.3, \langle k\rangle=320$.}
\label{Kplot}
\end{figure}

\subsection{Scaling of the algorithm}
Two types of scaling are commonly analyzed in high performance computing: the 
so-called ``weak'' scaling, which defines how the the time to solution varies 
when the resources are increased commensurately with the problem size, and the ``strong'' 
scaling, which is defined as how the time to solution decreases with an 
increasing amount of resources for fixed problem size.

We begin by analyzing the scaling of the time to solution for fixed resources 
but varying problem size. As ``problem size'' we consider the average 
expansion order or matrix size, $\langle k\rangle$. 
The average expansion 
order is related to the potential energy and therefore extensive in 
cluster size.
%In the absence of a ``sign 
%problem'' (sign problems will not be discussed in this article) this quantity 
For systems with small average expansion orders ($N\lesssim 200$), the entire 
matrix fits into the cache, and therefore there is no advantage  in using 
submatrix updates. With increasing average matrix size
caching effects become more important. 

Figure~\ref{SecUpdate} shows the strong scaling, as the time per update (in 
arbitrary units) as a function of the expansion order (matrix size), for rank-one 
updates and several $k_\text{max}$. 
The ideal scaling is $O(k^2)$ per update, or $O(k^3)$ for $\langle k\rangle$ 
updates needed to decorrelate a configuration.\footnote{In the presence of a sign problem there is an additional dependence of observable estimates on the average sign of the expansion -- we will not consider this case here.} 
The scaling per update is 
indicated by the dashed line.

Submatrix updates are, for problems with expansion orders between $512$ and $2048$, about a factor of eight faster than straightforward rank one updates.

For small expansion order CT-AUX with and without submatrix updates behave 
similarly.
For expansion orders of $256$ and larger, the speed increase from submatrix 
updates becomes apparent, and at expansion orders of $512$ and larger the 
difference with and without submatrix updates
corresponds to the difference of data transfer rates between the cache and CPU and 
the main memory and CPU, or the difference at which memory intensive (Sherman -
Morrison-like vector operations) and CPU intensive (Woodbury-like matrix 
operations) run.

The optimal choice of the expansion parameter $k_\text{max}$ for the test 
architecture lies somewhere between 64 and 128 (performance is relatively 
insensitive to the exact choice of $k_\text{max}$).
This is also illustrated in Fig.~\ref{Kplot}: for a small choice of 
$k_\text{max}$ the Woodbury matrix-matrix operations
do not dominate the calculation and the algorithm is similar to CT-AUX, where 
much time is spent idling at memory bottlenecks. Caching effects get more 
advantageous for larger $k_\text{max}$, until for $k_\text{max} \gtrsim 128$ 
most of the time is spent updating and inverting the $\Gamma$ matrices. Note, 
however, that the optimal value of $k_\text{max}$ is expected to depend on
architectural details such as the size of the cache.  

\begin{figure}[tb]
\includegraphics[width=0.8\columnwidth]{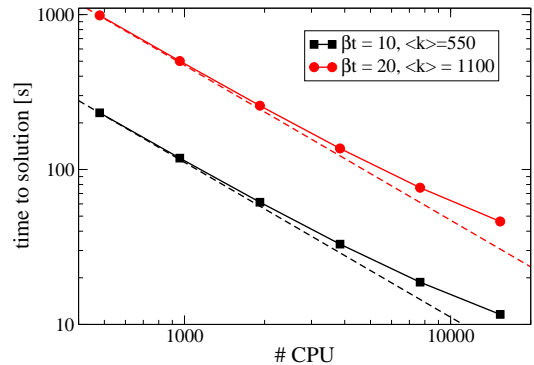}
\caption{Time to solution as a function of the number of CPUs, for a $16$-site cluster impurity problem. Squares (black online): $U/t=8$, $\beta t=10$ ($\langle k\rangle=550$), half filling. Circles (red online): $\beta t=20$ ($\langle k\rangle = 1100$). The dashed lines show the ideal scaling.}
\label{Nplot}
\end{figure}

In Fig.~\ref{Nplot} we present a strong scaling curve by showing the time to solution (in seconds) for two problem sets (symbols), as well as the ideal scaling (dashed lines), as a function of the number of CPUs employed.
This time includes communications and thermalization overhead that does not 
scale with the number of processors. This is the part that according to 
Amdahl's law\cite{Amdahl68} leads to less than ideal scaling behavior. CT-AUX 
has a remarkably small thermalization time and is therefore ideally suited 
for parallelization on large machines. As can be seen, for the chosen problem 
sizes, the algorithm can be scaled almost ideally to at least 10,000 CPUs.  
Note, however, that the scaling behavior is expected to depend critically on 
the number and type of measurements that are performed. This is because the 
measurements are perfectly parallel, since they are 
only performed once the calculation is thermalized.   Here, we have restricted 
the measurements to the single-particle Green's function.  If, in 
addition, more complex quantities such as two-particle observables are 
measured, the simulation run-time will be dominated by the measurements and 
the ideal scaling behavior is expected to continue to much larger processor 
counts.  

\begin{figure}[tb]
%{\includegraphics[width=0.95\columnwidth]{scale_U8_035.eps}}\\
{\includegraphics[width=0.95\columnwidth]{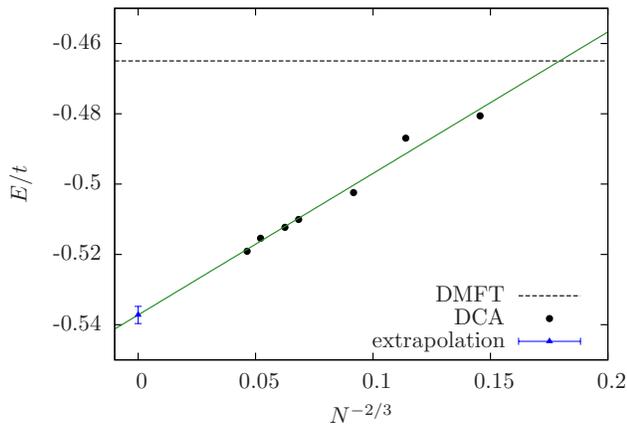}}
\caption{Extrapolation of the cluster energy as a function of cluster size, at $U/t = 8$, %$T/t = 0.35$ (upper panel) and 
$T/t=0.5$% (lower panel)
. Dashed line: DMFT results. Circles (black online) denote DCA results from clusters with size
$18$, $26,$ $36,$ $56,$ $64,$ $84,$ and  $100$. Solid line (green online): least squares fit. Triangle (blue online): extrapolated result. The error bar
denotes the fitting error; statistical (Monte Carlo) errors are smaller than symbol size.}
\label{Energy_extrapolation}
\end{figure}

\begin{figure*}[tb]
\begin{center}
%\subfigure[]{\label{selfT1}\includegraphics[width=0.3\textwidth]{self1.eps}}
%\subfigure[]{\label{selfT05}\includegraphics[width=0.3\textwidth]{self05.eps}}
%\subfigure[]{\label{selfT035}\includegraphics[width=0.3\textwidth]{self035.eps}}
\includegraphics{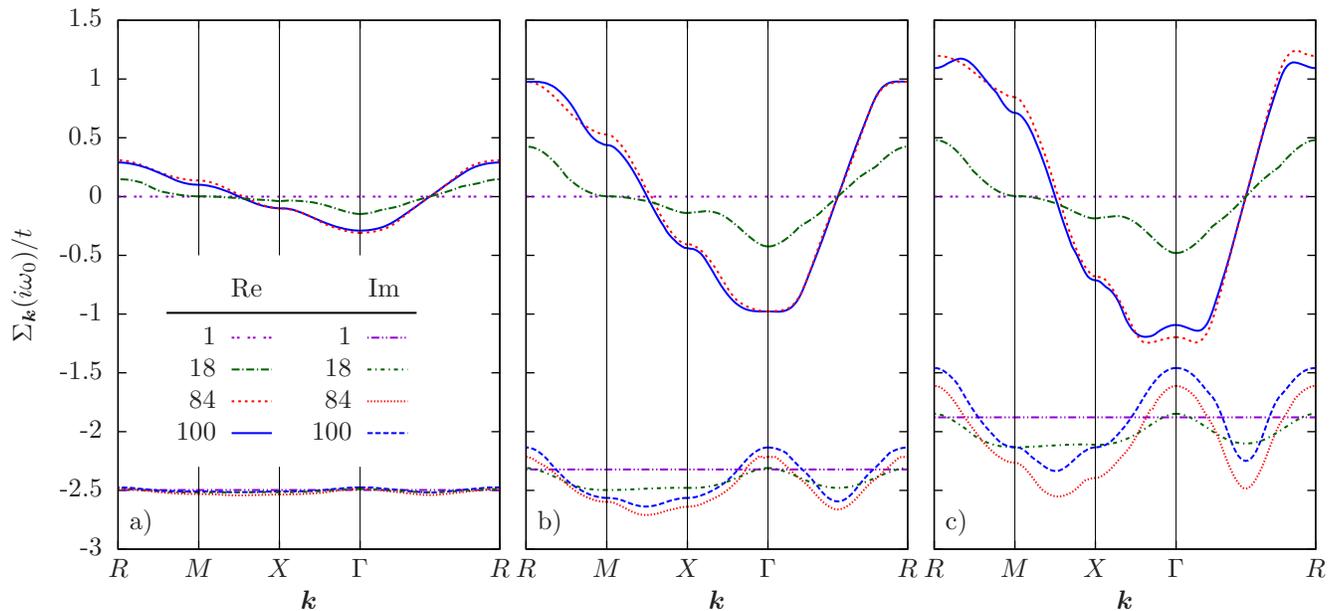}
\end{center}
\caption{Real and imaginary parts of the lowest Matsubara frequency of
  the interpolated DCA cluster self-energy $\Sigma(k, i\omega_0)$ of a
  $3D$ Hubbard model above the N\'eel temperature\cite{Fuchs10}, for
  $U/t=8$, $T/t=1$ (left panel), $T/t=0.5$ (middle panel), and
  $T/t=0.35$ (right panel), at half filling. The lines denote DMFT
  results (horizontal straight lines) and results for clusters of size
  $18$, $84$, and $100$. The interpolation follows a path along the
  high-symmetry points $\Gamma=(0,0,0)$, $X=(\pi,0,0)$,
  $M=(\pi,\pi,0)$, and $R=(\pi,\pi,\pi)$. }
\label{Selfenergy}
\end{figure*}

\begin{figure*}[tb]
\begin{center}
%\subfigure[]{\label{freqT05}\includegraphics[width=0.3\textwidth]{freqT05.eps}}
%\subfigure[]{\label{freqT05_pi}\includegraphics[width=0.3\textwidth]{freqT05_pi.eps}}
%\subfigure[]{\label{freqT05_05pi}\includegraphics[width=0.3\textwidth]{freqT05_05pi.eps}}\\
%\subfigure[]{\label{freqT035}\includegraphics[width=0.3\textwidth]{freqT035.eps}}
%\subfigure[]{\label{freqT035_pi}\includegraphics[width=0.3\textwidth]{freqT035_pi.eps}}
%\subfigure[]{\label{freqT035_05pi}\includegraphics[width=0.3\textwidth]{freqT035_05pi.eps}}
\includegraphics{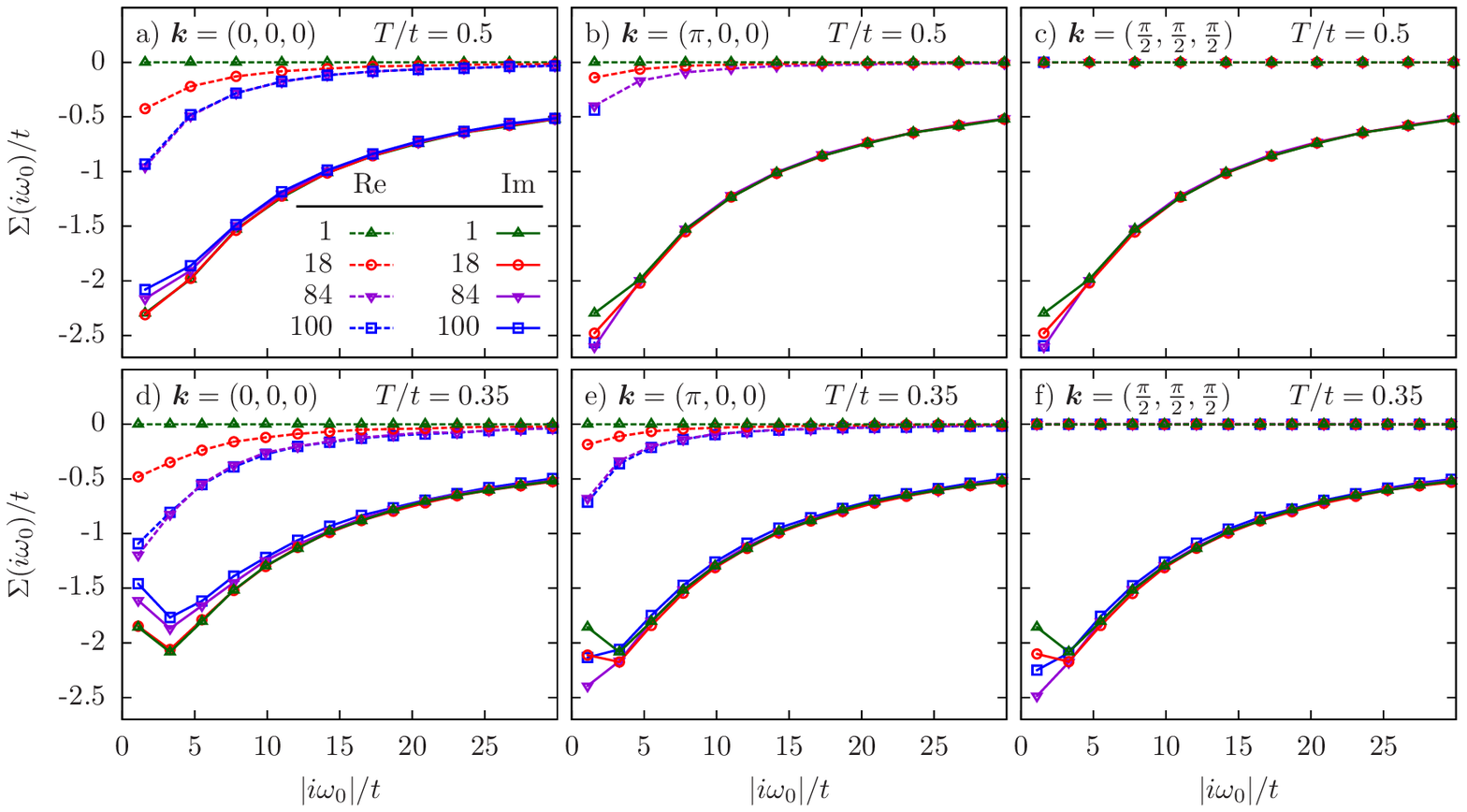}
\end{center}
\caption{Real and imaginary parts of the frequency dependence of the interpolated DCA cluster self-energy $\Sigma(k, i\omega_0)$ at selected $k$-points $(0,0,0)$, $(\pi,0,0)$, and $(\pi/2,\pi/2,\pi/2)$. $3D$ Hubbard model above the N\'eel temperature\cite{Fuchs10}, for $U/t=8$, $T/t=0.5$ (upper row), and $T/t=0.35$ (lower row), at half filling. The lines denote DMFT results (horizontal straight lines) and results for clusters of size $18$, $84$, and $100$.}
\label{Selfenergy_freq}
\end{figure*}

\subsection{Simulations of the $3D$ Hubbard model} As an illustration of the
power of the algorithm we present results from a calculation of the N\'{e}el
temperature of the three-dimensional Hubbard model at half filling, within the
DCA approximation, as a function of cluster size. 

A comprehensive study, showing DCA data at and away from half filling, for
interaction strengths up to $U \simeq 12$ and clusters of size $\leq$ $64$, will be published
elsewhere\cite{Fuchs10}. The results we present here are for temperature $T=t$ (far above $T_N$), for $T=0.5t$, and for $T=0.35t$.
The lowest temperature is close to the N\'{e}el temperature, and long ranged correlations cause a slow convergence. The results were obtained on 
$128$ CPUs in one hour per iteration. In Fig.~\ref{Energy_extrapolation} we show
the extrapolation of the energy for several cluster sizes and an extrapolation
to the infinite cluster size limit. The plot shows that controlled extrapolations to
the thermodynamic limit \cite{Maier05_dwave,Kent05,Kozik10,Fuchs10} can be
obtained in practice. Monte Carlo errors are much smaller than the symbol size.

Fig.~\ref{Selfenergy} shows self-energy cuts along the main axes in reciprocal
space. Plotted are results for single site DMFT and clusters of size $18$, $84$,
and $100$, interpolated using Akima splines. While momentum averaged
quantities like the energy in Fig.~\ref{Energy_extrapolation} show clear
convergence and the possibility for extrapolation, convergence is not uniform
in all quantities.  The high temperature self-energy plotted in panel~\ref{Selfenergy}a is clearly converged as a function of cluster size, the
intermediate temperature self-energy plotted in panel~\ref{Selfenergy}b shows some cluster size dependence, and the right panel~\ref{Selfenergy}c
shows a self-energy that even for $100$ cluster sites is not yet converged (a sign of the long wavelength physics important near $T_N$). Reliable extrapolation of the cluster self-energy to the
$\Sigma(k, \omega)$ of the infinite system would require even larger clusters.
Further insight can be gained from the frequency dependence of the Matsubara self-energy (Fig.~\ref{Selfenergy_freq}). Plotted is the frequency dependence at three points in the Brillouin zone.
While a significant cluster size dependence is observed at low frequencies, the results converge to the local DMFT limit at high frequencies, as one would expect.

\section{Conclusions}\label{concsec} We have presented a variation of the
CT-AUX algorithm that, while mathematically equivalent, arranges operations in
such a manner that they are ideally suited for modern computational
architectures. For large problem sizes, this ``submatrix'' algorithm achieves
a significant performance increase relative to the traditional CT-AUX
algorithm, by replacing the slow rank-one updates by faster matrix-matrix
operations.   Our implementation of the submatrix updates in the CT-AUX
algorithm requires an additional preparation step in which non-interacting
vertices with auxiliary spins $s=0$ are introduced.   After this step, the
CT-AUX vertex insertion and removal updates become equivalent to spin-flip
updates.   The submatrix algorithm then proceeds by manipulating the inverse
of the Green's function matrix, for which changes under auxiliary spin flips
are completely local.   This allows for a significantly faster computation of
the QMC transition probabilities under a spin-flip update.   The algorithm
keeps track of a number $k$ of these local changes, similar to the delayed
update algorithm, and then performs a Green's function update as
a matrix-matrix multiplication.   

Because this algorithm requires additional overhead over the traditional CT-AUX
implementation, there is an optimal choice for the maximum number of spin-flip
updates $k_\text{max}$ per Green's function update which depends on problem
size and  architectural parameters such as the cache size.  For the test
architecture we have used, we have found that $k_\text{max}\approx 128$ for 
large problem sizes.  For this optimal value, we find a speed increase up to a 
factor of 8 relative to the traditional CT-AUX algorithm.  

We have shown that simulations for large interacting systems, previously
requiring access to high performance supercomputers, become feasible for small
cluster architectures, and we have demonstrated the scaling on supercomputers
that shows that, by using the submatrix algorithm, continuous-time quantum
Monte Carlo methods are almost ideally adapted to high performance machines.

As an example we have shown how some cluster dynamical mean field theory
quantities, like the energy, can be reliably extrapolated to the thermodynamic
limit, and how for other quantities, like the self-energy, even large cluster
calculations are not sufficient to obtain converged extrapolations.

The algorithm is similarly suited to the solution of { lattice} problems [i.e., problems where $V_{p\sigma}=0$ and where no (cluster) dynamical mean field self-consistency is imposed]. 

Our results are also readily generalized to the interaction expansion formalism
developed in Refs.~\onlinecite{Rubtsov04,Rubtsov05}, offering the possibility to
significantly accelerate simulations of multi-orbital systems.

\acknowledgments{We acknowledge fruitful discussions with A. Lichtenstein, A. Millis, O. Parcollet, L. Pollet, M. Troyer, A. Georges, and P. Werner. The implementation of the submatrix updates is
based on the ALPS\cite{ALPS} library. Preliminary calculations were done on the
Brutus cluster at ETH Zurich. $3D$ calculations\cite{Fuchs10} used additional resources provided by GWDG and HLRN. Scaling calculations were performed on Jaguar at ORNL.
EG acknowledges funding by NSF DMR-0705847, SF and TP funding by the Deutsche Forschungsgemeinschaft through SFB 602.
This research used resources of the Oak Ridge Leadership Computing Facility at the Oak Ridge National Laboratory, which is supported by the Office of Science of the U.S. Department of Energy under Contract No. DE-AC05-00OR22725.
The research was conducted at the Center for Nanophase Materials
Sciences, which is sponsored at Oak Ridge National Laboratory by the Division
of Scientific User Facilities, U.S. Department of Energy, under project
CNMS2009-219.  } \bibliography{refs_shortened} 
\begin{thebibliography}{48}%
\makeatletter
\providecommand \@ifxundefined [1]{%
\@ifx{#1\undefined}
}%
\providecommand \@ifnum [1]{%
\ifnum #1\expandafter \@firstoftwo
\else \expandafter \@secondoftwo
\fi
}%
\providecommand \@ifx [1]{%
\ifx #1\expandafter \@firstoftwo
\else \expandafter \@secondoftwo
\fi
}%
\providecommand \natexlab [1]{#1}%
\providecommand \enquote  [1]{``#1''}%
\providecommand \bibnamefont  [1]{#1}%
\providecommand \bibfnamefont [1]{#1}%
\providecommand \citenamefont [1]{#1}%
\providecommand \href@noop [0]{\@secondoftwo}%
\providecommand \href [0]{\begingroup \@sanitize@url \@href}%
\providecommand \@href[1]{\@@startlink{#1}\@@href}%
\providecommand \@@href[1]{\endgroup#1\@@endlink}%
\providecommand \@sanitize@url [0]{\catcode `\\12\catcode `\$12\catcode
`\&12\catcode `\#12\catcode `\^12\catcode `\_12\catcode `\%12\relax}%
\providecommand \@@startlink[1]{}%
\providecommand \@@endlink[0]{}%
\providecommand \url  [0]{\begingroup\@sanitize@url \@url }%
\providecommand \@url [1]{\endgroup\@href {#1}{\urlprefix }}%
\providecommand \urlprefix  [0]{URL }%
\providecommand \Eprint [0]{\href }%
\providecommand \doibase [0]{http://dx.doi.org/}%
\providecommand \selectlanguage [0]{\@gobble}%
\providecommand \bibinfo  [0]{\@secondoftwo}%
\providecommand \bibfield  [0]{\@secondoftwo}%
\providecommand \translation [1]{[#1]}%
\providecommand \BibitemOpen [0]{}%
\providecommand \bibitemStop [0]{}%
\providecommand \bibitemNoStop [0]{.\EOS\space}%
\providecommand \EOS [0]{\spacefactor3000\relax}%
\providecommand \BibitemShut  [1]{\csname bibitem#1\endcsname}%
\let\auto@bib@innerbib\@empty
%</preamble>
\bibitem [{\citenamefont {Dagotto}(1991)}]{Dagotto91}%
\BibitemOpen
\bibfield  {author} {\bibinfo {author} {\bibfnamefont {E.}~\bibnamefont
{Dagotto}},\ }\href {\doibase 10.1142/S0217979291000067} {\bibfield
{journal} {\bibinfo  {journal} {Int. J. Mod. Phys. B}\ }\textbf {\bibinfo
{volume} {5}},\ \bibinfo {pages} {77} (\bibinfo {year} {1991})}\BibitemShut
{NoStop}%
\bibitem [{\citenamefont {White}(1992)}]{White92}%
\BibitemOpen
\bibfield  {author} {\bibinfo {author} {\bibfnamefont {S.~R.}\ \bibnamefont
{White}},\ }\href {\doibase 10.1103/PhysRevLett.69.2863} {\bibfield
{journal} {\bibinfo  {journal} {Phys. Rev. Lett.}\ }\textbf {\bibinfo
{volume} {69}},\ \bibinfo {pages} {2863} (\bibinfo {year}
{1992})}\BibitemShut {NoStop}%
\bibitem [{\citenamefont {Schollw\"{o}ck}(2005)}]{Schollwoeck05}%
\BibitemOpen
\bibfield  {author} {\bibinfo {author} {\bibfnamefont {U.}~\bibnamefont
{Schollw\"{o}ck}},\ }\href {\doibase 10.1103/RevModPhys.77.259} {\bibfield
{journal} {\bibinfo  {journal} {Rev. Mod. Phys.}\ }\textbf {\bibinfo {volume}
{77}},\ \bibinfo {eid} {259} (\bibinfo {year} {2005})}\BibitemShut {NoStop}%
\bibitem [{\citenamefont {Verstraete}\ and\ \citenamefont
{Cirac}(2004)}]{PEPS}%
\BibitemOpen
\bibfield  {author} {\bibinfo {author} {\bibfnamefont {F.}~\bibnamefont
{Verstraete}}\ and\ \bibinfo {author} {\bibfnamefont {J.}~\bibnamefont
{Cirac}},\ }\href@noop {} {\enquote {\bibinfo {title} {Renormalization
algorithms for quantum-many body systems in two and higher dimensions},}\ }
(\bibinfo {year} {2004}),\ \Eprint {http://arxiv.org/abs/cond-mat/0407066v1}
{arXiv:cond-mat/0407066v1} \BibitemShut {NoStop}%
\bibitem [{\citenamefont {Vidal}(2007)}]{Vidal05}%
\BibitemOpen
\bibfield  {author} {\bibinfo {author} {\bibfnamefont {G.}~\bibnamefont
{Vidal}},\ }\href {\doibase 10.1103/PhysRevLett.99.220405} {\bibfield
{journal} {\bibinfo  {journal} {Phys. Rev. Lett.}\ }\textbf {\bibinfo
{volume} {99}},\ \bibinfo {eid} {220405} (\bibinfo {year}
{2007})}\BibitemShut {NoStop}%
\bibitem [{\citenamefont {Schuch}\ \emph {et~al.}(2008)\citenamefont {Schuch},
\citenamefont {Wolf}, \citenamefont {Verstraete},\ and\ \citenamefont
{Cirac}}]{Schuch08}%
\BibitemOpen
\bibfield  {author} {\bibinfo {author} {\bibfnamefont {N.}~\bibnamefont
{Schuch}}, \bibinfo {author} {\bibfnamefont {M.~M.}\ \bibnamefont {Wolf}},
\bibinfo {author} {\bibfnamefont {F.}~\bibnamefont {Verstraete}}, \ and\
\bibinfo {author} {\bibfnamefont {J.~I.}\ \bibnamefont {Cirac}},\ }\href
{\doibase 10.1103/PhysRevLett.100.040501} {\bibfield  {journal} {\bibinfo
{journal} {Phys. Rev. Lett.}\ }\textbf {\bibinfo {volume} {100}},\ \bibinfo
{eid} {040501} (\bibinfo {year} {2008})}\BibitemShut {NoStop}%
\bibitem [{\citenamefont {Corboz}\ \emph {et~al.}(2010)\citenamefont {Corboz},
\citenamefont {Evenbly}, \citenamefont {Verstraete},\ and\ \citenamefont
{Vidal}}]{Corboz10}%
\BibitemOpen
\bibfield  {author} {\bibinfo {author} {\bibfnamefont {P.}~\bibnamefont
{Corboz}}, \bibinfo {author} {\bibfnamefont {G.}~\bibnamefont {Evenbly}},
\bibinfo {author} {\bibfnamefont {F.}~\bibnamefont {Verstraete}}, \ and\
\bibinfo {author} {\bibfnamefont {G.}~\bibnamefont {Vidal}},\ }\href
{\doibase 10.1103/PhysRevA.81.010303} {\bibfield  {journal} {\bibinfo
{journal} {Phys. Rev. A}\ }\textbf {\bibinfo {volume} {81}},\ \bibinfo
{pages} {010303} (\bibinfo {year} {2010})}\BibitemShut {NoStop}%
\bibitem [{\citenamefont {Blankenbecler}\ \emph {et~al.}(1981)\citenamefont
{Blankenbecler}, \citenamefont {Scalapino},\ and\ \citenamefont
{Sugar}}]{BSS81}%
\BibitemOpen
\bibfield  {author} {\bibinfo {author} {\bibfnamefont {R.}~\bibnamefont
{Blankenbecler}}, \bibinfo {author} {\bibfnamefont {D.~J.}\ \bibnamefont
{Scalapino}}, \ and\ \bibinfo {author} {\bibfnamefont {R.~L.}\ \bibnamefont
{Sugar}},\ }\href {\doibase 10.1103/PhysRevD.24.2278} {\bibfield  {journal}
{\bibinfo  {journal} {Phys. Rev. D}\ }\textbf {\bibinfo {volume} {24}},\
\bibinfo {pages} {2278} (\bibinfo {year} {1981})}\BibitemShut {NoStop}%
\bibitem [{\citenamefont {Loh}\ \emph {et~al.}(1990)\citenamefont {Loh},
\citenamefont {Gubernatis}, \citenamefont {Scalettar}, \citenamefont {White},
\citenamefont {Scalapino},\ and\ \citenamefont {Sugar}}]{Loh90}%
\BibitemOpen
\bibfield  {author} {\bibinfo {author} {\bibfnamefont {E.~Y.}\ \bibnamefont
{Loh}}, \bibinfo {author} {\bibfnamefont {J.~E.}\ \bibnamefont {Gubernatis}},
\bibinfo {author} {\bibfnamefont {R.~T.}\ \bibnamefont {Scalettar}}, \bibinfo
{author} {\bibfnamefont {S.~R.}\ \bibnamefont {White}}, \bibinfo {author}
{\bibfnamefont {D.~J.}\ \bibnamefont {Scalapino}}, \ and\ \bibinfo {author}
{\bibfnamefont {R.~L.}\ \bibnamefont {Sugar}},\ }\href {\doibase 10.1103/PhysRevB.41.9301} {\bibfield  {journal} {\bibinfo  {journal} {Phys.
Rev. B}\ }\textbf {\bibinfo {volume} {41}},\ \bibinfo {pages} {9301}
(\bibinfo {year} {1990})}\BibitemShut {NoStop}%
\bibitem [{\citenamefont {Troyer}\ and\ \citenamefont
{Wiese}(2005)}]{Troyer05}%
\BibitemOpen
\bibfield  {author} {\bibinfo {author} {\bibfnamefont {M.}~\bibnamefont
{Troyer}}\ and\ \bibinfo {author} {\bibfnamefont {U.-J.}\ \bibnamefont
{Wiese}},\ }\href {\doibase 10.1103/PhysRevLett.94.170201} {\bibfield
{journal} {\bibinfo  {journal} {Phys. Rev. Lett.}\ }\textbf {\bibinfo
{volume} {94}},\ \bibinfo {eid} {170201} (\bibinfo {year}
{2005})}\BibitemShut {NoStop}%
\bibitem [{\citenamefont {Georges}\ \emph {et~al.}(1996)\citenamefont
{Georges}, \citenamefont {Kotliar}, \citenamefont {Krauth},\ and\
\citenamefont {Rozenberg}}]{Georges96}%
\BibitemOpen
\bibfield  {author} {\bibinfo {author} {\bibfnamefont {A.}~\bibnamefont
{Georges}}, \bibinfo {author} {\bibfnamefont {G.}~\bibnamefont {Kotliar}},
\bibinfo {author} {\bibfnamefont {W.}~\bibnamefont {Krauth}}, \ and\ \bibinfo
{author} {\bibfnamefont {M.~J.}\ \bibnamefont {Rozenberg}},\ }\href {\doibase 10.1103/RevModPhys.68.13} {\bibfield  {journal} {\bibinfo  {journal} {Rev.
Mod. Phys.}\ }\textbf {\bibinfo {volume} {68}},\ \bibinfo {pages} {13}
(\bibinfo {year} {1996})}\BibitemShut {NoStop}%
\bibitem [{\citenamefont {Kotliar}\ \emph {et~al.}(2006)\citenamefont
{Kotliar}, \citenamefont {Savrasov}, \citenamefont {Haule} \emph
{et~al.}}]{Kotliar06}%
\BibitemOpen
\bibfield  {author} {\bibinfo {author} {\bibfnamefont {G.}~\bibnamefont
{Kotliar}}, \bibinfo {author} {\bibfnamefont {S.~Y.}\ \bibnamefont
{Savrasov}}, \bibinfo {author} {\bibfnamefont {K.}~\bibnamefont {Haule}},
\emph {et~al.},\ }\href {\doibase 10.1103/RevModPhys.78.865} {\bibfield
{journal} {\bibinfo  {journal} {Rev. Mod. Phys.}\ }\textbf {\bibinfo {volume}
{78}},\ \bibinfo {eid} {865} (\bibinfo {year} {2006})}\BibitemShut {NoStop}%
\bibitem [{\citenamefont {Metzner}\ and\ \citenamefont
{Vollhardt}(1989)}]{Metzner89}%
\BibitemOpen
\bibfield  {author} {\bibinfo {author} {\bibfnamefont {W.}~\bibnamefont
{Metzner}}\ and\ \bibinfo {author} {\bibfnamefont {D.}~\bibnamefont
{Vollhardt}},\ }\href {\doibase 10.1103/PhysRevLett.62.324} {\bibfield
{journal} {\bibinfo  {journal} {Phys. Rev. Lett.}\ }\textbf {\bibinfo
{volume} {62}},\ \bibinfo {pages} {324} (\bibinfo {year} {1989})}\BibitemShut
{NoStop}%
\bibitem [{\citenamefont {M\"{u}ller-Hartmann}(1989)}]{MH89}%
\BibitemOpen
\bibfield  {author} {\bibinfo {author} {\bibfnamefont {E.}~\bibnamefont
{M\"{u}ller-Hartmann}},\ }\href {\doibase 10.1007/BF01311397} {\bibfield
{journal} {\bibinfo  {journal} {Z. Phys. B}\ }\textbf {\bibinfo {volume}
{74}},\ \bibinfo {pages} {507} (\bibinfo {year} {1989})}\BibitemShut
{NoStop}%
\bibitem [{\citenamefont {Georges}\ and\ \citenamefont
{Kotliar}(1992)}]{Georges92b}%
\BibitemOpen
\bibfield  {author} {\bibinfo {author} {\bibfnamefont {A.}~\bibnamefont
{Georges}}\ and\ \bibinfo {author} {\bibfnamefont {G.}~\bibnamefont
{Kotliar}},\ }\href {\doibase 10.1103/PhysRevB.45.6479} {\bibfield  {journal}
{\bibinfo  {journal} {Phys. Rev. B}\ }\textbf {\bibinfo {volume} {45}},\
\bibinfo {pages} {6479} (\bibinfo {year} {1992})}\BibitemShut {NoStop}%
\bibitem [{\citenamefont {Hettler}\ \emph {et~al.}(1998)\citenamefont
{Hettler}, \citenamefont {Tahvildar-Zadeh}, \citenamefont {Jarrell} \emph
{et~al.}}]{Hettler98}%
\BibitemOpen
\bibfield  {author} {\bibinfo {author} {\bibfnamefont {M.~H.}\ \bibnamefont
{Hettler}}, \bibinfo {author} {\bibfnamefont {A.~N.}\ \bibnamefont
{Tahvildar-Zadeh}}, \bibinfo {author} {\bibfnamefont {M.}~\bibnamefont
{Jarrell}},  \emph {et~al.},\ }\href {\doibase 10.1103/PhysRevB.58.R7475}
{\bibfield  {journal} {\bibinfo  {journal} {Phys. Rev. B}\ }\textbf {\bibinfo
{volume} {58}},\ \bibinfo {pages} {R7475} (\bibinfo {year}
{1998})}\BibitemShut {NoStop}%
\bibitem [{\citenamefont {Lichtenstein}\ and\ \citenamefont
{Katsnelson}(2000)}]{Lichtenstein00}%
\BibitemOpen
\bibfield  {author} {\bibinfo {author} {\bibfnamefont {A.~I.}\ \bibnamefont
{Lichtenstein}}\ and\ \bibinfo {author} {\bibfnamefont {M.~I.}\ \bibnamefont
{Katsnelson}},\ }\href {\doibase 10.1103/PhysRevB.62.R9283} {\bibfield
{journal} {\bibinfo  {journal} {Phys. Rev. B}\ }\textbf {\bibinfo {volume}
{62}},\ \bibinfo {pages} {R9283} (\bibinfo {year} {2000})}\BibitemShut
{NoStop}%
\bibitem [{\citenamefont {Hettler}\ \emph {et~al.}(2000)\citenamefont
{Hettler}, \citenamefont {Mukherjee}, \citenamefont {Jarrell},\ and\
\citenamefont {Krishnamurthy}}]{Hettler00}%
\BibitemOpen
\bibfield  {author} {\bibinfo {author} {\bibfnamefont {M.~H.}\ \bibnamefont
{Hettler}}, \bibinfo {author} {\bibfnamefont {M.}~\bibnamefont {Mukherjee}},
\bibinfo {author} {\bibfnamefont {M.}~\bibnamefont {Jarrell}}, \ and\
\bibinfo {author} {\bibfnamefont {H.~R.}\ \bibnamefont {Krishnamurthy}},\
}\href {\doibase 10.1103/PhysRevB.61.12739} {\bibfield  {journal} {\bibinfo
{journal} {Phys. Rev. B}\ }\textbf {\bibinfo {volume} {61}},\ \bibinfo
{pages} {12739} (\bibinfo {year} {2000})}\BibitemShut {NoStop}%
\bibitem [{\citenamefont {Kotliar}\ \emph {et~al.}(2001)\citenamefont
{Kotliar}, \citenamefont {Savrasov}, \citenamefont {P\'alsson},\ and\
\citenamefont {Biroli}}]{Kotliar01}%
\BibitemOpen
\bibfield  {author} {\bibinfo {author} {\bibfnamefont {G.}~\bibnamefont
{Kotliar}}, \bibinfo {author} {\bibfnamefont {S.~Y.}\ \bibnamefont
{Savrasov}}, \bibinfo {author} {\bibfnamefont {G.}~\bibnamefont {P\'alsson}},
\ and\ \bibinfo {author} {\bibfnamefont {G.}~\bibnamefont {Biroli}},\ }\href
{\doibase 10.1103/PhysRevLett.87.186401} {\bibfield  {journal} {\bibinfo
{journal} {Phys. Rev. Lett.}\ }\textbf {\bibinfo {volume} {87}},\ \bibinfo
{pages} {186401} (\bibinfo {year} {2001})}\BibitemShut {NoStop}%
\bibitem [{\citenamefont {Maier}\ \emph
{et~al.}(2005{\natexlab{a}})\citenamefont {Maier}, \citenamefont {Jarrell},
\citenamefont {Pruschke},\ and\ \citenamefont {Hettler}}]{Maier05}%
\BibitemOpen
\bibfield  {author} {\bibinfo {author} {\bibfnamefont {T.}~\bibnamefont
{Maier}}, \bibinfo {author} {\bibfnamefont {M.}~\bibnamefont {Jarrell}},
\bibinfo {author} {\bibfnamefont {T.}~\bibnamefont {Pruschke}}, \ and\
\bibinfo {author} {\bibfnamefont {M.~H.}\ \bibnamefont {Hettler}},\ }\href
{\doibase 10.1103/RevModPhys.77.1027} {\bibfield  {journal} {\bibinfo
{journal} {Rev. Mod. Phys.}\ }\textbf {\bibinfo {volume} {77}},\ \bibinfo
{eid} {1027} (\bibinfo {year} {2005}{\natexlab{a}})}\BibitemShut {NoStop}%
\bibitem [{\citenamefont {Okamoto}\ \emph {et~al.}(2003)\citenamefont
{Okamoto}, \citenamefont {Millis}, \citenamefont {Monien},\ and\
\citenamefont {Fuhrmann}}]{Okamoto03}%
\BibitemOpen
\bibfield  {author} {\bibinfo {author} {\bibfnamefont {S.}~\bibnamefont
{Okamoto}}, \bibinfo {author} {\bibfnamefont {A.~J.}\ \bibnamefont {Millis}},
\bibinfo {author} {\bibfnamefont {H.}~\bibnamefont {Monien}}, \ and\ \bibinfo
{author} {\bibfnamefont {A.}~\bibnamefont {Fuhrmann}},\ }\href {\doibase 10.1103/PhysRevB.68.195121} {\bibfield  {journal} {\bibinfo  {journal} {Phys.
Rev. B}\ }\textbf {\bibinfo {volume} {68}},\ \bibinfo {pages} {195121}
(\bibinfo {year} {2003})}\BibitemShut {NoStop}%
\bibitem [{\citenamefont {Bulla}\ \emph {et~al.}(1998)\citenamefont {Bulla},
\citenamefont {Hewson},\ and\ \citenamefont {Pruschke}}]{Bulla08}%
\BibitemOpen
\bibfield  {author} {\bibinfo {author} {\bibfnamefont {R.}~\bibnamefont
{Bulla}}, \bibinfo {author} {\bibfnamefont {A.~C.}\ \bibnamefont {Hewson}}, \
and\ \bibinfo {author} {\bibfnamefont {T.}~\bibnamefont {Pruschke}},\ }\href
{\doibase 10.1088/0953-8984/10/37/021} {\bibfield  {journal} {\bibinfo
{journal} {J. Phys. Condens. Matter}\ }\textbf {\bibinfo {volume} {10}},\
\bibinfo {pages} {8365} (\bibinfo {year} {1998})}\BibitemShut {NoStop}%
\bibitem [{\citenamefont {Caffarel}\ and\ \citenamefont
{Krauth}(1994)}]{Caffarel94}%
\BibitemOpen
\bibfield  {author} {\bibinfo {author} {\bibfnamefont {M.}~\bibnamefont
{Caffarel}}\ and\ \bibinfo {author} {\bibfnamefont {W.}~\bibnamefont
{Krauth}},\ }\href {\doibase 10.1103/PhysRevLett.72.1545} {\bibfield
{journal} {\bibinfo  {journal} {Phys. Rev. Lett.}\ }\textbf {\bibinfo
{volume} {72}},\ \bibinfo {pages} {1545} (\bibinfo {year}
{1994})}\BibitemShut {NoStop}%
\bibitem [{\citenamefont {Pruschke}\ and\ \citenamefont
{Grewe}(1989)}]{Pruschke89}%
\BibitemOpen
\bibfield  {author} {\bibinfo {author} {\bibfnamefont {T.}~\bibnamefont
{Pruschke}}\ and\ \bibinfo {author} {\bibfnamefont {N.}~\bibnamefont
{Grewe}},\ }\href {\doibase 10.1007/BF01311391} {\bibfield  {journal}
{\bibinfo  {journal} {Z. Phys. B}\ }\textbf {\bibinfo {volume} {74}},\
\bibinfo {pages} {439} (\bibinfo {year} {1989})}\BibitemShut {NoStop}%
\bibitem [{\citenamefont {Coleman}(1984)}]{Coleman84}%
\BibitemOpen
\bibfield  {author} {\bibinfo {author} {\bibfnamefont {P.}~\bibnamefont
{Coleman}},\ }\href {\doibase 10.1103/PhysRevB.29.3035} {\bibfield  {journal}
{\bibinfo  {journal} {Phys. Rev. B}\ }\textbf {\bibinfo {volume} {29}},\
\bibinfo {pages} {3035} (\bibinfo {year} {1984})}\BibitemShut {NoStop}%
\bibitem [{\citenamefont {Hirsch}\ and\ \citenamefont {Fye}(1986)}]{Hirsch86}%
\BibitemOpen
\bibfield  {author} {\bibinfo {author} {\bibfnamefont {J.~E.}\ \bibnamefont
{Hirsch}}\ and\ \bibinfo {author} {\bibfnamefont {R.~M.}\ \bibnamefont
{Fye}},\ }\href {\doibase 10.1103/PhysRevLett.56.2521} {\bibfield  {journal}
{\bibinfo  {journal} {Phys. Rev. Lett.}\ }\textbf {\bibinfo {volume} {56}},\
\bibinfo {pages} {2521} (\bibinfo {year} {1986})}\BibitemShut {NoStop}%
\bibitem [{\citenamefont {Rubtsov}\ and\ \citenamefont
{Lichtenstein}(2004)}]{Rubtsov04}%
\BibitemOpen
\bibfield  {author} {\bibinfo {author} {\bibfnamefont {A.~N.}\ \bibnamefont
{Rubtsov}}\ and\ \bibinfo {author} {\bibfnamefont {A.~I.}\ \bibnamefont
{Lichtenstein}},\ }\href {\doibase 10.1134/1.1800216} {\bibfield  {journal}
{\bibinfo  {journal} {JETP Letters}\ }\textbf {\bibinfo {volume} {80}},\
\bibinfo {pages} {61} (\bibinfo {year} {2004})}\BibitemShut {NoStop}%
\bibitem [{\citenamefont {Rubtsov}\ \emph {et~al.}(2005)\citenamefont
{Rubtsov}, \citenamefont {Savkin},\ and\ \citenamefont
{Lichtenstein}}]{Rubtsov05}%
\BibitemOpen
\bibfield  {author} {\bibinfo {author} {\bibfnamefont {A.~N.}\ \bibnamefont
{Rubtsov}}, \bibinfo {author} {\bibfnamefont {V.~V.}\ \bibnamefont {Savkin}},
\ and\ \bibinfo {author} {\bibfnamefont {A.~I.}\ \bibnamefont
{Lichtenstein}},\ }\href {\doibase 10.1103/PhysRevB.72.035122} {\bibfield
{journal} {\bibinfo  {journal} {Phys. Rev. B}\ }\textbf {\bibinfo {volume}
{72}},\ \bibinfo {eid} {035122} (\bibinfo {year} {2005})}\BibitemShut
{NoStop}%
\bibitem [{\citenamefont {Werner}\ \emph {et~al.}(2006)\citenamefont {Werner},
\citenamefont {Comanac}, \citenamefont {de' Medici} \emph
{et~al.}}]{Werner06}%
\BibitemOpen
\bibfield  {author} {\bibinfo {author} {\bibfnamefont {P.}~\bibnamefont
{Werner}}, \bibinfo {author} {\bibfnamefont {A.}~\bibnamefont {Comanac}},
\bibinfo {author} {\bibfnamefont {L.}~\bibnamefont {de' Medici}},  \emph
{et~al.},\ }\href {\doibase 10.1103/PhysRevLett.97.076405} {\bibfield
{journal} {\bibinfo  {journal} {Phys. Rev. Lett.}\ }\textbf {\bibinfo
{volume} {97}},\ \bibinfo {eid} {076405} (\bibinfo {year}
{2006})}\BibitemShut {NoStop}%
\bibitem [{\citenamefont {Werner}\ and\ \citenamefont
{Millis}(2006)}]{Werner06Kondo}%
\BibitemOpen
\bibfield  {author} {\bibinfo {author} {\bibfnamefont {P.}~\bibnamefont
{Werner}}\ and\ \bibinfo {author} {\bibfnamefont {A.~J.}\ \bibnamefont
{Millis}},\ }\href {\doibase 10.1103/PhysRevB.74.155107} {\bibfield
{journal} {\bibinfo  {journal} {Phys. Rev. B}\ }\textbf {\bibinfo {volume}
{74}},\ \bibinfo {eid} {155107} (\bibinfo {year} {2006})}\BibitemShut
{NoStop}%
\bibitem [{\citenamefont {Gull}\ \emph {et~al.}(2008)\citenamefont {Gull},
\citenamefont {Werner}, \citenamefont {Parcollet},\ and\ \citenamefont
{Troyer}}]{Gull08_ctaux}%
\BibitemOpen
\bibfield  {author} {\bibinfo {author} {\bibfnamefont {E.}~\bibnamefont
{Gull}}, \bibinfo {author} {\bibfnamefont {P.}~\bibnamefont {Werner}},
\bibinfo {author} {\bibfnamefont {O.}~\bibnamefont {Parcollet}}, \ and\
\bibinfo {author} {\bibfnamefont {M.}~\bibnamefont {Troyer}},\ }\href
{\doibase 10.1209/0295-5075/82/57003} {\bibfield  {journal} {\bibinfo
{journal} {Europhys. Lett.}\ }\textbf {\bibinfo {volume} {82}},\ \bibinfo
{pages} {57003 (6pp)} (\bibinfo {year} {2008})}\BibitemShut {NoStop}%
\bibitem [{\citenamefont {L\"auchli}\ and\ \citenamefont
{Werner}(2009)}]{laeuchli09}%
\BibitemOpen
\bibfield  {author} {\bibinfo {author} {\bibfnamefont {A.~M.}\ \bibnamefont
{L\"auchli}}\ and\ \bibinfo {author} {\bibfnamefont {P.}~\bibnamefont
{Werner}},\ }\href {\doibase 10.1103/PhysRevB.80.235117} {\bibfield
{journal} {\bibinfo  {journal} {Phys. Rev. B}\ }\textbf {\bibinfo {volume}
{80}},\ \bibinfo {pages} {235117} (\bibinfo {year} {2009})}\BibitemShut
{NoStop}%
\bibitem [{\citenamefont {Gull}\ \emph {et~al.}(2007)\citenamefont {Gull},
\citenamefont {Werner}, \citenamefont {Millis},\ and\ \citenamefont
{Troyer}}]{Gull07}%
\BibitemOpen
\bibfield  {author} {\bibinfo {author} {\bibfnamefont {E.}~\bibnamefont
{Gull}}, \bibinfo {author} {\bibfnamefont {P.}~\bibnamefont {Werner}},
\bibinfo {author} {\bibfnamefont {A.}~\bibnamefont {Millis}}, \ and\ \bibinfo
{author} {\bibfnamefont {M.}~\bibnamefont {Troyer}},\ }\href {\doibase 10.1103/PhysRevB.76.235123} {\bibfield  {journal} {\bibinfo  {journal} {Phys.
Rev. B}\ }\textbf {\bibinfo {volume} {76}},\ \bibinfo {eid} {235123}
(\bibinfo {year} {2007})}\BibitemShut {NoStop}%
\bibitem [{\citenamefont {Alvarez}\ \emph {et~al.}(2008)\citenamefont
{Alvarez}, \citenamefont {Summers}, \citenamefont {Maxwell}, \citenamefont
{Eisenbach}, \citenamefont {Meredith}, \citenamefont {Larkin}, \citenamefont
{Levesque}, \citenamefont {Maier}, \citenamefont {Kent}, \citenamefont
{D'Azevedo},\ and\ \citenamefont {Schulthess}}]{Alvarez08}%
\BibitemOpen
\bibfield  {author} {\bibinfo {author} {\bibfnamefont {G.}~\bibnamefont
{Alvarez}}, \bibinfo {author} {\bibfnamefont {M.~S.}\ \bibnamefont
{Summers}}, \bibinfo {author} {\bibfnamefont {D.~E.}\ \bibnamefont
{Maxwell}}, \bibinfo {author} {\bibfnamefont {M.}~\bibnamefont {Eisenbach}},
\bibinfo {author} {\bibfnamefont {J.~S.}\ \bibnamefont {Meredith}}, \bibinfo
{author} {\bibfnamefont {J.~M.}\ \bibnamefont {Larkin}}, \bibinfo {author}
{\bibfnamefont {J.}~\bibnamefont {Levesque}}, \bibinfo {author}
{\bibfnamefont {T.~A.}\ \bibnamefont {Maier}}, \bibinfo {author}
{\bibfnamefont {P.~R.~C.}\ \bibnamefont {Kent}}, \bibinfo {author}
{\bibfnamefont {E.~F.}\ \bibnamefont {D'Azevedo}}, \ and\ \bibinfo {author}
{\bibfnamefont {T.~C.}\ \bibnamefont {Schulthess}},\ }in\ \href@noop {}
{\emph {\bibinfo {booktitle} {SC '08: Proceedings of the 2008 ACM/IEEE
conference on Supercomputing}}}\ (\bibinfo  {publisher} {IEEE Press},\
\bibinfo {address} {Piscataway, NJ, USA},\ \bibinfo {year} {2008})\ pp.\
\bibinfo {pages} {1--10}\BibitemShut {NoStop}%
\bibitem [{\citenamefont {Nukala}\ \emph {et~al.}(2009)\citenamefont {Nukala},
\citenamefont {Maier}, \citenamefont {Summers}, \citenamefont {Alvarez},\
and\ \citenamefont {Schulthess}}]{Nukala09}%
\BibitemOpen
\bibfield  {author} {\bibinfo {author} {\bibfnamefont {P.~K. V.~V.}\
\bibnamefont {Nukala}}, \bibinfo {author} {\bibfnamefont {T.~A.}\
\bibnamefont {Maier}}, \bibinfo {author} {\bibfnamefont {M.~S.}\ \bibnamefont
{Summers}}, \bibinfo {author} {\bibfnamefont {G.}~\bibnamefont {Alvarez}}, \
and\ \bibinfo {author} {\bibfnamefont {T.~C.}\ \bibnamefont {Schulthess}},\
}\href {\doibase 10.1103/PhysRevB.80.195111} {\bibfield  {journal} {\bibinfo
{journal} {Phys. Rev. B}\ }\textbf {\bibinfo {volume} {80}},\ \bibinfo {eid}
{195111} (\bibinfo {year} {2009})}\BibitemShut {NoStop}%
\bibitem [{\citenamefont {Werner}\ \emph {et~al.}(2009)\citenamefont {Werner},
\citenamefont {Gull}, \citenamefont {Parcollet},\ and\ \citenamefont
{Millis}}]{Werner09_8site}%
\BibitemOpen
\bibfield  {author} {\bibinfo {author} {\bibfnamefont {P.}~\bibnamefont
{Werner}}, \bibinfo {author} {\bibfnamefont {E.}~\bibnamefont {Gull}},
\bibinfo {author} {\bibfnamefont {O.}~\bibnamefont {Parcollet}}, \ and\
\bibinfo {author} {\bibfnamefont {A.~J.}\ \bibnamefont {Millis}},\ }\href
{\doibase 10.1103/PhysRevB.80.045120} {\bibfield  {journal} {\bibinfo
{journal} {Phys. Rev. B}\ }\textbf {\bibinfo {volume} {80}},\ \bibinfo
{pages} {045120} (\bibinfo {year} {2009})}\BibitemShut {NoStop}%
\bibitem [{\citenamefont {Mikelsons}(2009)}]{KarlisThesis}%
\BibitemOpen
\bibfield  {author} {\bibinfo {author} {\bibfnamefont {K.}~\bibnamefont
{Mikelsons}},\ }\href@noop {} {Ph.D. thesis},\ \bibinfo  {school} {University
of Cincinnati} (\bibinfo {year} {2009})\BibitemShut {NoStop}%
\bibitem [{\citenamefont {Rombouts}\ \emph {et~al.}(1999)\citenamefont
{Rombouts}, \citenamefont {Heyde},\ and\ \citenamefont
{Jachowicz}}]{Rombouts99}%
\BibitemOpen
\bibfield  {author} {\bibinfo {author} {\bibfnamefont {S.~M.~A.}\
\bibnamefont {Rombouts}}, \bibinfo {author} {\bibfnamefont {K.}~\bibnamefont
{Heyde}}, \ and\ \bibinfo {author} {\bibfnamefont {N.}~\bibnamefont
{Jachowicz}},\ }\href {\doibase 10.1103/PhysRevLett.82.4155} {\bibfield
{journal} {\bibinfo  {journal} {Phys. Rev. Lett.}\ }\textbf {\bibinfo
{volume} {82}},\ \bibinfo {pages} {4155} (\bibinfo {year}
{1999})}\BibitemShut {NoStop}%
\bibitem [{\citenamefont {Werner}\ \emph {et~al.}(2010)\citenamefont {Werner},
\citenamefont {Oka}, \citenamefont {Eckstein},\ and\ \citenamefont
{Millis}}]{Werner10}%
\BibitemOpen
\bibfield  {author} {\bibinfo {author} {\bibfnamefont {P.}~\bibnamefont
{Werner}}, \bibinfo {author} {\bibfnamefont {T.}~\bibnamefont {Oka}},
\bibinfo {author} {\bibfnamefont {M.}~\bibnamefont {Eckstein}}, \ and\
\bibinfo {author} {\bibfnamefont {A.~J.}\ \bibnamefont {Millis}},\ }\href
{\doibase 10.1103/PhysRevB.81.035108} {\bibfield  {journal} {\bibinfo
{journal} {Phys. Rev. B}\ }\textbf {\bibinfo {volume} {81}},\ \bibinfo
{pages} {035108} (\bibinfo {year} {2010})}\BibitemShut {NoStop}%
\bibitem [{\citenamefont {Sherman}\ and\ \citenamefont
{Morrison}(1950)}]{ShermanMorrison50}%
\BibitemOpen
\bibfield  {author} {\bibinfo {author} {\bibfnamefont {J.}~\bibnamefont
{Sherman}}\ and\ \bibinfo {author} {\bibfnamefont {W.~J.}\ \bibnamefont
{Morrison}},\ }\href@noop {} {\bibfield  {journal} {\bibinfo  {journal} {The
Annals of Mathematical Statistics}\ }\textbf {\bibinfo {volume} {21}},\
\bibinfo {pages} {124} (\bibinfo {year} {1950})}\BibitemShut {NoStop}%
\bibitem [{\citenamefont {Bennett}(1965)}]{Bennett65}%
\BibitemOpen
\bibfield  {author} {\bibinfo {author} {\bibfnamefont {J.~M.}\ \bibnamefont
{Bennett}},\ }\href {\doibase 10.1007/BF01436076} {\bibfield  {journal}
{\bibinfo  {journal} {Numerische Mathematik}\ }\textbf {\bibinfo {volume}
{7}},\ \bibinfo {pages} {217} (\bibinfo {year} {1965})}\BibitemShut {NoStop}%
\bibitem [{Note1()}]{Note1}%
\BibitemOpen
\bibinfo {note} {In the presence of a sign problem there is an additional
dependence of observable estimates on the average sign of the expansion -- we
will not consider this case here.}\BibitemShut {Stop}%
\bibitem [{\citenamefont {Amdahl}(1967)}]{Amdahl68}%
\BibitemOpen
\bibfield  {author} {\bibinfo {author} {\bibfnamefont {G.}~\bibnamefont
{Amdahl}},\ }in\ \href@noop {} {\emph {\bibinfo {booktitle} {AFIPS Conference
Proceedings}}},\ Vol.~\bibinfo {volume} {30}\ (\bibinfo {year} {1967})\ pp.\
\bibinfo {pages} {483 -- 485}\BibitemShut {NoStop}%
\bibitem [{\citenamefont {Fuchs}\ \emph {et~al.}(2010)\citenamefont {Fuchs},
\citenamefont {Gull}, \citenamefont {Pollet}, \citenamefont {Burovski},
\citenamefont {Kozik}, \citenamefont {Pruschke},\ and\ \citenamefont
{Troyer}}]{Fuchs10}%
\BibitemOpen
\bibfield  {author} {\bibinfo {author} {\bibfnamefont {S.}~\bibnamefont
{Fuchs}}, \bibinfo {author} {\bibfnamefont {E.}~\bibnamefont {Gull}},
\bibinfo {author} {\bibfnamefont {L.}~\bibnamefont {Pollet}}, \bibinfo
{author} {\bibfnamefont {E.}~\bibnamefont {Burovski}}, \bibinfo {author}
{\bibfnamefont {E.}~\bibnamefont {Kozik}}, \bibinfo {author} {\bibfnamefont
{T.}~\bibnamefont {Pruschke}}, \ and\ \bibinfo {author} {\bibfnamefont
{M.}~\bibnamefont {Troyer}},\ }\href@noop {} {} (\bibinfo {year} {2010}),\
\Eprint {http://arxiv.org/abs/1009.2759} {arXiv:1009.2759} \BibitemShut
{NoStop}%
\bibitem [{\citenamefont {Maier}\ \emph
{et~al.}(2005{\natexlab{b}})\citenamefont {Maier}, \citenamefont {Jarrell},
\citenamefont {Schulthess}, \citenamefont {Kent},\ and\ \citenamefont
{White}}]{Maier05_dwave}%
\BibitemOpen
\bibfield  {author} {\bibinfo {author} {\bibfnamefont {T.~A.}\ \bibnamefont
{Maier}}, \bibinfo {author} {\bibfnamefont {M.}~\bibnamefont {Jarrell}},
\bibinfo {author} {\bibfnamefont {T.~C.}\ \bibnamefont {Schulthess}},
\bibinfo {author} {\bibfnamefont {P.~R.~C.}\ \bibnamefont {Kent}}, \ and\
\bibinfo {author} {\bibfnamefont {J.~B.}\ \bibnamefont {White}},\ }\href
{\doibase 10.1103/PhysRevLett.95.237001} {\bibfield  {journal} {\bibinfo
{journal} {Phys. Rev. Lett.}\ }\textbf {\bibinfo {volume} {95}},\ \bibinfo
{pages} {237001} (\bibinfo {year} {2005}{\natexlab{b}})}\BibitemShut
{NoStop}%
\bibitem [{\citenamefont {Kent}\ \emph {et~al.}(2005)\citenamefont {Kent},
\citenamefont {Jarrell}, \citenamefont {Maier},\ and\ \citenamefont
{Pruschke}}]{Kent05}%
\BibitemOpen
\bibfield  {author} {\bibinfo {author} {\bibfnamefont {P.~R.~C.}\
\bibnamefont {Kent}}, \bibinfo {author} {\bibfnamefont {M.}~\bibnamefont
{Jarrell}}, \bibinfo {author} {\bibfnamefont {T.~A.}\ \bibnamefont {Maier}},
\ and\ \bibinfo {author} {\bibfnamefont {T.}~\bibnamefont {Pruschke}},\
}\href {\doibase 10.1103/PhysRevB.72.060411} {\bibfield  {journal} {\bibinfo
{journal} {Phys. Rev. B}\ }\textbf {\bibinfo {volume} {72}},\ \bibinfo
{pages} {060411} (\bibinfo {year} {2005})}\BibitemShut {NoStop}%
\bibitem [{\citenamefont {Kozik}\ \emph {et~al.}(2010)\citenamefont {Kozik},
\citenamefont {Houcke}, \citenamefont {Gull}, \citenamefont {Pollet},
\citenamefont {Prokof'ev}, \citenamefont {Svistunov},\ and\ \citenamefont
{Troyer}}]{Kozik10}%
\BibitemOpen
\bibfield  {author} {\bibinfo {author} {\bibfnamefont {E.}~\bibnamefont
{Kozik}}, \bibinfo {author} {\bibfnamefont {K.~V.}\ \bibnamefont {Houcke}},
\bibinfo {author} {\bibfnamefont {E.}~\bibnamefont {Gull}}, \bibinfo {author}
{\bibfnamefont {L.}~\bibnamefont {Pollet}}, \bibinfo {author} {\bibfnamefont
{N.}~\bibnamefont {Prokof'ev}}, \bibinfo {author} {\bibfnamefont
{B.}~\bibnamefont {Svistunov}}, \ and\ \bibinfo {author} {\bibfnamefont
{M.}~\bibnamefont {Troyer}},\ }\href {\doibase 10.1209/0295-5075/90/10004}
{\bibfield  {journal} {\bibinfo  {journal} {Europhys. Lett.}\ }\textbf
{\bibinfo {volume} {90}},\ \bibinfo {pages} {10004} (\bibinfo {year}
{2010})}\BibitemShut {NoStop}%
\bibitem [{\citenamefont {Albuquerque}\ \emph {et~al.}(2007)\citenamefont
{Albuquerque}, \citenamefont {Alet}, \citenamefont {Corboz} \emph
{et~al.}}]{ALPS}%
\BibitemOpen
\bibfield  {author} {\bibinfo {author} {\bibfnamefont {A.}~\bibnamefont
{Albuquerque}}, \bibinfo {author} {\bibfnamefont {F.}~\bibnamefont {Alet}},
\bibinfo {author} {\bibfnamefont {P.}~\bibnamefont {Corboz}},  \emph
{et~al.},\ }\href {\doibase 10.1016/j.jmmm.2006.10.304} {\bibfield  {journal}
{\bibinfo  {journal} {J. Magn. Magn. Mater.}\ }\textbf {\bibinfo {volume}
{310}},\ \bibinfo {pages} {1187} (\bibinfo {year} {2007})}\BibitemShut
{NoStop}%
\end{thebibliography}%
\end{document}